\documentclass[iop,apj,tighten]{emulateapj}

\shorttitle{CN AND CH$^+$ IN DIFFUSE MOLECULAR CLOUDS}
\shortauthors{RITCHEY, FEDERMAN, \& LAMBERT}

\begin{document}
\title{Interstellar CN and CH$^+$ in Diffuse Molecular Clouds: $^{12}$C/$^{13}$C 
Ratios and CN Excitation\altaffilmark{1}}
\author{A. M. Ritchey\altaffilmark{2}$^,$\altaffilmark{3}, S. R. 
Federman\altaffilmark{2}$^,$\altaffilmark{3}, and D. L. Lambert\altaffilmark{4}}
\altaffiltext{1}{Based in part on observations made with the Very Large 
Telescope of the European Southern Observatory, Paranal, Chile, under programs 
065.I-0526, 071.C-0367, 071.C-0513, and 076.C-0431.}
\altaffiltext{2}{Department of Physics and Astronomy, University of Toledo, 
Toledo, OH 43606; adam.ritchey@utoledo.edu; steven.federman@utoledo.edu.}
\altaffiltext{3}{Guest Observer, McDonald Observatory, University of Texas at 
Austin, Austin, TX 78712.}
\altaffiltext{4}{W. J. McDonald Observatory, University of Texas at Austin, 
Austin, TX 78712; dll@astro.as.utexas.edu.}

\begin{abstract}

We present very high signal-to-noise ratio absorption-line observations of CN 
and CH$^+$ along 13 lines of sight through diffuse molecular clouds. The data 
are examined to extract precise isotopologic ratios of $^{12}$CN/$^{13}$CN and 
$^{12}$CH$^+$/$^{13}$CH$^+$ in order to assess predictions of diffuse cloud 
chemistry. Our results on $^{12}$CH$^+$/$^{13}$CH$^+$ confirm that this ratio 
does not deviate from the ambient $^{12}$C/$^{13}$C ratio in local interstellar 
clouds, as expected if the formation of CH$^+$ involves nonthermal processes. 
We find that $^{12}$CN/$^{13}$CN, however, can be significantly fractionated 
away from the ambient value. The dispersion in our sample of $^{12}$CN/$^{13}$CN 
ratios is similar to that found in recent surveys of $^{12}$CO/$^{13}$CO. For 
sight lines where both ratios have been determined, the $^{12}$CN/$^{13}$CN 
ratios are generally fractionated in the opposite sense compared to 
$^{12}$CO/$^{13}$CO. Chemical fractionation in CO results from competition 
between selective photodissociation and isotopic charge exchange. An inverse 
relationship between $^{12}$CN/$^{13}$CN and $^{12}$CO/$^{13}$CO follows from the 
coexistence of CN and CO in diffuse cloud cores. However, an isotopic charge 
exchange reaction with CN may mitigate the enhancements in $^{12}$CN/$^{13}$CN 
for lines of sight with low $^{12}$CO/$^{13}$CO ratios. For two sight lines with 
high values of $^{12}$CO/$^{13}$CO, our results indicate that about 50\% of the 
carbon is locked up in CO, which is consistent with the notion that these sight 
lines probe molecular cloud envelopes where the transition from C$^+$ to CO is 
expected to occur. An analysis of CN rotational excitation yields a weighted 
mean value for $T_{01}$($^{12}$CN) of $2.754\pm0.002$ K, which implies an excess 
over the temperature of the cosmic microwave background of only $29\pm3$ mK. 
This modest excess eliminates the need for a local excitation mechanism beyond 
electron and neutral collisions. The rotational excitation temperatures in 
$^{13}$CN show no excess over the temperature of the CMB.

\end{abstract}


\section{INTRODUCTION}

The $^{12}$C-to-$^{13}$C isotopic ratio in the interstellar medium (ISM) is an 
important diagnostic for probing the history of nucleosynthesis and chemical 
enrichment in the Galaxy. The abundance of $^{13}$C increases relative to 
$^{12}$C over Galactic timescales because $^{13}$C is a secondary product of 
stellar nucleosynthesis. Moreover, millimeter-wave emission studies of CO 
(Langer \& Penzias 1990), H$_{2}$CO (Henkel et al. 1982), and CN (Savage et al. 
2002; Milam et al. 2005) in dense molecular clouds have demonstrated the 
existence of a Galactic gradient in the $^{12}$C/$^{13}$C ratio, with 
increasingly lower values found closer to the Galactic center (see also the 
reviews by Wilson \& Rood 1994 and Wilson 1999). The increase in 
$^{12}$C/$^{13}$C with Galactocentric distance presumably reflects the reduced 
rates of star formation, and subsequent stellar nucleosynthesis and mass loss, 
that are characteristic of the outer Galactic disk. The enrichment process, 
acting over the 4.6~Gyr since the formation of the solar system, also offers a 
natural explanation for the reduction in the $^{12}$C/$^{13}$C ratio in the 
vicinity of the Sun. The terrestrial value of $^{12}$C/$^{13}$C, which should 
approximate the interstellar value at the time of collapse of the solar nebula, 
is 89 (Lodders 2003), while values close to 70 are typical of local 
interstellar clouds. Estimates for the present-day, local ISM value of 
$^{12}$C/$^{13}$C are usually obtained from absorption or emission studies of 
carbon-bearing molecules, which serve as proxies when direct measurements of 
the ambient carbon isotopic ratio are lacking. Wilson (1999), for example, 
derived an average $^{12}$C/$^{13}$C ratio of $69\pm6$ from millimeter-wave data 
on CO and H$_{2}$CO emission for 13 local sources. More recently, Milam et al. 
(2005) added millimeter observations of CN in dense clouds to those of CO and 
H$_{2}$CO and found an average for the present-day, local $^{12}$C/$^{13}$C 
ratio of $68\pm15$. In this paper, we shall follow Sheffer et al. (2007) and 
take the ambient $^{12}$C/$^{13}$C ratio in the solar neighborhood to be 
$70\pm7$ (see also Sheffer et al. 2002).

When molecular proxies are used to evaluate the interstellar abundance ratio of 
$^{12}$C to $^{13}$C, the values sometimes show evidence of chemical 
fractionation. While this is generally not the case in dense molecular clouds, 
where the processes responsible for fractionation are less effective (see Milam 
et al. 2005), fractionation can have a significant impact in diffuse clouds. 
Detailed models predict that the fractionation processes in diffuse 
environments will not affect each interstellar molecule in the same way. The 
molecular ion CH$^+$ is not subject to chemical fractionation because it must 
be formed at high effective temperatures. The production of CH$^+$ is linked to 
the endothermic reaction C$^+$~+~H$_2$~$\to$~CH$^+$~+~H (Elitzur \& Watson 
1978, 1980), which has an activation energy of $\Delta E/k_{\mathrm{B}}=4640$ K. 
Since CH$^+$ is associated with cold diffuse clouds, its formation requires 
that nonthermal processes, such as magnetohydrodynamic shocks or propagating 
Alfv\'en waves, provide the additional heating. As a result, the 
$^{12}$CH$^+$/$^{13}$CH$^+$ ratio is believed to be equilibrated and thus to be 
the best measure of the ambient carbon isotopic ratio in the diffuse ISM. Many 
observations (e.g., Centuri\'on \& Vladilo 1991; Crane et al. 1991; Stahl \& 
Wilson 1992; Centuri\'on et al. 1995) have confirmed this expectation for 
CH$^+$, revealing $^{12}$CH$^+$/$^{13}$CH$^+$ ratios very near 70 for local 
diffuse clouds. Centuri\'on et al. (1995) give a weighted mean value of 
$^{12}$CH$^+$/$^{13}$CH$^+$ of $67\pm3$ for five sight lines to stars within 
approximately 500 pc of the Sun. However, some authors (e.g., Hawkins \& Jura 
1987; Vladilo et al. 1993; Casassus et al. 2005; Stahl et al. 2008) have 
reported discrepant values of $^{12}$CH$^+$/$^{13}$CH$^+$. Vladilo et al. (1993) 
found ratios of $126\pm29$ and $98\pm19$ toward HD~152235 and HD~152424, 
respectively. Since both of these stars belong to the Sco OB1 association (at 
$d\simeq2$~kpc), Vladilo et al. (1993) suggest that $^{12}$CH$^+$/$^{13}$CH$^+$ 
may vary on scales of $\sim$1 kpc or smaller, depending on the distance to the 
clouds. Hawkins \& Jura (1987) found $^{12}$CH$^+$/$^{13}$CH$^+$ ratios of 
$40\pm9$ and $41\pm9$ toward 20~Tau and 23~Tau, respectively. Because these 
stars are members of the nearby Pleiades cluster ($d=110$ pc), such small 
values of $^{12}$CH$^+$/$^{13}$CH$^+$ seem to challenge the view that a ratio 
near 70 characterizes the solar neighborhood.

Carbon-bearing molecules susceptible to fractionation, especially CO, may 
exhibit $^{12}$C-to-$^{13}$C ratios that are either enhanced or reduced with 
respect to the ambient value. Two competing processes are capable of altering 
the $^{12}$CO/$^{13}$CO ratio in diffuse molecular gas. Selective 
photodissociation (SPD) favors $^{12}$CO since it is the more abundant 
isotopologue and is thus protected to a greater extent via self shielding 
(e.g., van Dishoeck \& Black 1988; Visser et al. 2009). The process is 
effective in the case of CO because the photodissociation of this molecule is 
governed by line absorption. At lower gas kinetic temperatures, CO is 
influenced by the isotopic charge exchange (ICE) reaction 
$^{13}$C$^+$~+~$^{12}$CO~$\to$~$^{12}$C$^+$~+~$^{13}$CO~+~$\Delta E$, where the 
difference in zero-point energies $\Delta E/k_{\mathrm{B}}=35$~K (Watson et al. 
1976) favors $^{13}$CO. Because CO is the most abundant carbon-bearing molecule 
in the ISM, an enhancement in $^{13}$CO, resulting from isotopic charge 
exchange, will deplete the carbon reservoir of $^{13}$C. Conversely, if 
$^{13}$CO is selectively destroyed through photodissociation, then the carbon 
reservoir will be enhanced in $^{13}$C. Any molecule arising from the remaining 
carbon in the reservoir should therefore possess a $^{12}$C-to-$^{13}$C ratio 
that is fractionated in the opposite sense compared to the ratio in CO. Such 
behavior is expected for CN because this molecule coexists with CO in diffuse 
molecular clouds (Pan et al. 2005). Existing data on isotopologic ratios in CO, 
CN, and CH$^+$ for the well-studied sight line to $\zeta$~Oph offer some 
evidence that the chemical predictions are borne out. The diffuse gas in this 
direction has $^{12}$CO/$^{13}$CO = $167\pm15$ (Lambert et al. 1994), 
$^{12}$CN/$^{13}$CN = $47.3^{+5.5}_{-4.4}$ (Crane \& Hegyi 1988), and 
$^{12}$CH$^+$/$^{13}$CH$^+$ = $67.5\pm4.5$ (Crane et al. 1991). Thus, the ratios 
in CO and CN are fractionated in opposing directions, while the ratio in CH$^+$ 
is consistent with the presumed ambient carbon isotopic ratio.

In this investigation, we seek to measure $^{12}$CN/$^{13}$CN and 
$^{12}$CH$^+$/$^{13}$CH$^+$ ratios along lines of sight where $^{12}$CO/$^{13}$CO 
is either enhanced or reduced. Recent UV surveys of $^{12}$CO/$^{13}$CO along 
diffuse and translucent sight lines (Sonnentrucker et al. 2007; Burgh et al. 
2007; Sheffer et al. 2007) have typically yielded ratios consistent with the 
average value of $^{12}$C/$^{13}$C for local interstellar clouds. Indeed, the 
weighted mean value of $^{12}$CO/$^{13}$CO for the 25 sight lines studied by 
Sheffer et al. (2007) is $70\pm2$. However, there are some notable exceptions. 
Enhanced ratios are found, not only in the direction of $\zeta$~Oph, but also 
toward $\rho$~Oph~A and $\chi$~Oph, where the respective $^{12}$CO/$^{13}$CO 
ratios are $125\pm23$ and $117\pm35$ (Federman et al. 2003), as well as toward 
$\zeta$~Per, where $^{12}$CO/$^{13}$CO = $108\pm5$ (Sheffer et al. 2007). 
Reduced ratios, with respect to the ambient value, are found in the directions 
of 20~Aql, where $^{12}$CO/$^{13}$CO = $50\pm15$ (Hanson et al. 1992), and 
HD~154368, where $^{12}$CO/$^{13}$CO = $37\pm8$ (Sheffer et al. 2007), among 
several others (see Sheffer et al. 2007). Sonnentrucker et al. (2007) find a 
low $^{12}$CO/$^{13}$CO ratio toward HD~73882 ($25\pm22$), but the relative 
uncertainties are large. Our goal is to obtain $^{12}$CN/$^{13}$CN and 
$^{12}$CH$^+$/$^{13}$CH$^+$ ratios for as many of these sight lines as possible, 
so that, when our results are combined with the existing data on 
$^{12}$CO/$^{13}$CO, the suite of measurements can be used to test the 
predictions of chemical models for diffuse clouds. We reexamine the sight line 
to $\zeta$~Oph as a check on our general methodology. We also obtain 
$^{12}$CH$^+$/$^{13}$CH$^+$ ratios toward the Pleiades stars, 20~Tau and 23~Tau, 
to investigate potential scatter in the carbon isotopic ratio within the solar 
neighborhood.

Since high-quality data are needed to detect the weak features associated with 
$^{13}$CN and $^{13}$CH$^+$, the observations examined here also allow precise 
determinations of rotational excitation temperatures in CN. In some situations, 
these measurements can then be used to constrain the electron density, an 
important parameter for modeling the physical conditions within a given cloud 
(e.g., Black \& van Dishoeck 1991). It is widely understood that the CN 
molecule is maintained in radiative equilibrium with the cosmic microwave 
background (CMB) in interstellar space (see the review by Thaddeus 1972). The 
CMB is the primary source of radiation in the universe at 2.64 mm and 1.32 mm, 
the wavelengths of the two lowest rotational transitions in CN. As a result, CN 
excitation temperatures reflect the temperature of the CMB at these 
wavelengths, in the absence of any local sources of excitation. Observed 
excitation temperatures in CN, in fact, do exhibit an excess over the 
temperature of the CMB, as derived from the Far Infrared Absolute 
Spectrophotometer (FIRAS) onboard the \emph{COBE} satellite 
($T_{\mathrm{CMB}}=2.725\pm0.002$ K; Mather et al. 1999; see also Fixsen 2009), 
but this excess is small ($<0.1$ K; e.g., Palazzi et al. 1992). Since electron 
impact should dominate any local contribution to CN excitation (Thaddeus 1972), 
an observed excess provides an estimate for the density of electrons in the 
portion of the cloud traced by CN.

In order to investigate isotopologic ratios in CN and CH$^+$ and CN rotational 
excitation in diffuse molecular clouds, we examine high-resolution, very high 
signal-to-noise ratio observations of optical absorption lines arising from 
electronic transitions within the CN $B$~$^2\Sigma^+-$$X$~$^2\Sigma^+$ and 
CH$^+$ $A$ $^1\Pi-$$X$ $^1\Sigma^+$ systems. The observations and data 
reduction procedures are described in \S{} 2. In \S{}~3, we provide detailed 
information concerning the profile synthesis routine, with which we derive our 
final column densities and isotopologic ratios. The analysis and discussion of 
our results on $^{12}$CH$^+$/$^{13}$CH$^+$ and $^{12}$CN/$^{13}$CN ratios appears 
in \S\S{} 4.1 and 4.2, respectively. In \S{} 4.3, we explore the relationship 
between $^{12}$CN/$^{13}$CN and $^{12}$CO/$^{13}$CO in an effort to evaluate the 
effects of chemical fractionation in diffuse molecular gas. The topic of CN 
rotational excitation is examined in \S{} 5, and our main findings are 
summarized in \S{} 6. An appendix gives results on weak Ca~{\small I} and 
Ca~{\small II} absorption toward $\alpha$ Leo and $\alpha$ Vir, which were 
observed as unreddened standard stars to aid in the reduction of the McDonald 
Observatory data (see \S{} 2.1).

\section{OBSERVATIONS AND DATA REDUCTION}

\subsection{McDonald Observatory Spectra}
Principal data for this project were acquired with the Harlan J. Smith 2.7 m 
telescope at McDonald Observatory using the Tull (2dcoud\'e) spectrograph (Tull 
et al. 1995) in its high-resolution mode (cs21). The observations were carried 
out over the course of four observing runs between 2007 January and 2008 May. 
Two winter runs (2007 January and 2007 December/2008 January) provided data on 
20~Tau, 23~Tau, and $\zeta$~Per, while spectra of $\rho$~Oph~A, $\zeta$~Oph, 
and 20~Aql were acquired during two summer observing sessions (2007 June and 
2008 May). For all of the observations, the crossed-dispersed echelle 
spectrometer was configured with the 79 gr mm$^{-1}$ grating (E1), the 145 
$\mu$m slit (Slit 2), and a $2048\times2048$ CCD (TK3). Order 56 of the E1 
grating spectrum was centered at 4077 \AA, enabling both the CN $B$$-$$X$ 
(0,~0) lines near 3874 \AA{} and the CH$^+$ $A$$-$$X$ (0,~0) transition at 4232 
\AA{} to be recorded with a single exposure\footnote{This setting also provides 
data on the CH $A$$-$$X$ (0,~0) transition at 4300 \AA{} and on the 
Ca~{\scriptsize I} $\lambda$4226 and Ca~{\scriptsize II} $\lambda$3933 lines.}. 
The grating tilt was adjusted slightly each night so that the interstellar 
features would not always fall on the same portion of the CCD. This helps to 
mitigate the effect of fixed pattern noise and reduces the chance that an 
instrumental glitch at a specific location on the CCD will negatively impact 
the observations.

The detection of absorption from the $^{13}$C-bearing isotopologues of CN and 
CH$^+$ requires spectra with very high signal-to-noise ratios (S/N $>1000$ per 
resolution element). To achieve this, the program stars were observed for 
several hours each night. Individual stellar exposures, taken in succession, 
were limited to 30 minutes to minimize the number of cosmic ray hits in each
integration. A bright, unreddened star was observed nightly to the same signal 
level as the program stars to confirm that no instrumental defects were present 
at the locations of interstellar features. For both of the winter observing 
runs, $\alpha$ Leo (Regulus; B7 V; $B=1.24$) served as the unreddened standard 
star. For the summer observations, either $\alpha$ Vir (Spica; B1 III-IV+; 
$B=0.91$) or, on one occasion, $\alpha$ Lyr (Vega; A0 V; $B=0.03$) was observed 
for this purpose. Calibration exposures for dark current and bias correction, 
flat fielding, and wavelength assignment were obtained during each of the 
observing runs. Four 30-minute dark frames were acquired on the first night of 
a run, while a series of ten biases and ten flats were taken each night. For 
accurate wavelength calibration, Th-Ar comparison spectra were recorded 
throughout the night at intervals of 2$-$3 hr.

Reduction of the raw McDonald data employed standard routines within the Image 
Reduction and Analysis Facility (IRAF) environment. First, the overscan region 
in each of the raw images was fitted with a low-order polynomial and the excess 
counts were removed from the exposure. An average bias was created for each 
night and was used to correct the darks, flats, stellar exposures, and 
comparison lamp frames. No dark correction was applied to any of the 
observations because the level of dark current was always found to be 
insignificant after the bias was removed. Cosmic rays were eliminated from the 
stellar and comparison lamp exposures by setting the respective thresholds for 
detection to 50 and 25 times the mean of the surrounding pixels. Any cosmic 
rays present in individual flat lamp exposures were effectively removed by 
taking the median of all flats for a given night. Scattered light was modeled 
in the dispersion and cross-dispersion directions and subtracted from the 
stellar exposures and from the median flat. The flat was then normalized to 
unity and divided into the stellar and comparison lamp frames to account for 
(first-order) pixel-to-pixel variations across the CCD. One-dimensional spectra 
were extracted from the processed images by summing across the width of each 
order without weighting the individual pixels.

At this point, the extracted spectra of the program stars were examined in 
pixel space to determine the pixel coordinates of the interstellar absorption 
lines. The combined spectrum of the unreddened star from each night was then 
inspected at these positions to determine whether any instrumental defects 
remained after flat fielding. No obvious glitches were found in the vicinity of 
CN or CH$^+$. However, low level fixed pattern noise was identified in some of 
the stellar spectra, particularly those with the highest S/N. To correct for 
these second-order fluctuations, the affected stellar exposures were divided by 
the normalized spectrum of the unreddened star for that night. The same 
technique, when applied to exposures with lower S/N, did not significantly 
improve the quality of the data and so was not adopted for these exposures. 
Next, stellar spectra were calibrated in wavelength after identifying emission 
lines in the Th-Ar comparison spectra, typically 5 per order. The dispersion 
solution applied to a given stellar exposure resulted from an interpolation 
between the solutions derived from the Th-Ar lamp exposures that preceded and 
followed the stellar exposure. Finally, the calibrated spectra were shifted to 
the reference frame of the local standard of rest (LSR).

All exposures of a given target from the multiple observing runs were coadded 
to maximize the S/N in the final spectrum. Before inclusion in the sum, each 
exposure was carefully examined, specifically near the expected positions of 
the $^{13}$CN and $^{13}$CH$^+$ absorption features, to determine if any weak 
cosmic rays had survived the reduction process. When cosmic rays were 
identified, they were removed manually if they occurred far enough away from 
the interstellar features. If the cosmic ray affected an interstellar feature 
directly, then the exposure was removed from the sum. Because the CN and CH$^+$ 
lines reside in different echelle orders and the orders were coadded 
separately, an exposure removed from the sum in one order would not necessarily 
be removed in the other. Thus, the total exposure times for the two spectral 
regions may differ somewhat (see Table 1). Further complicating the coaddition 
process was a slight change in the resolving power of the spectrograph from one 
run to the next. The measured widths of thorium emission lines in the Th-Ar 
comparison spectra indicate a resolving power of $R=173,000$ ($\Delta$$v=1.7$ 
km s$^{-1}$) was achieved during our run in 2007 January. This decreased to 
150,000 (2.0 km s$^{-1}$) for the run in June of that year, but returned to 
160,000 (1.9 km s$^{-1}$) and 168,000 (1.8~km~s$^{-1}$) for the runs in 2007 
December and 2008 May, respectively. When combining exposures from multiple 
observing runs, the data were sampled at the dispersion of the lowest 
resolution spectra contributing to the sum.

Final coadded spectra were normalized to the continuum by fitting low-order 
polynomials to regions free of interstellar absorption within a narrow spectral 
window (no more than 2 \AA{} wide) surrounding a given interstellar line. The 
immediate vicinity of the CN and CH$^+$ lines in our McDonald spectra are free 
of contamination by either telluric or instrumental features, making continuum 
normalization straightforward. In only one case did the placement of the 
continuum present any significant difficulties. The interstellar $^{12}$CH$^+$ 
line toward 20~Tau is superimposed onto a narrow photospheric Fe~{\small II} 
feature at 4233 \AA{}. This situation is unique among the McDonald targets to 
20~Tau, which has a relatively late spectral type and a rather slow rotation 
rate ($v$ sin $i=39$ km s$^{-1}$; Hoffleit \& Jaschek 1982). Figure 1 shows a 
range of possible continuum fits to the regions surrounding the CH$^+$ lines 
toward 20~Tau. The small variation in the continuum above the $^{13}$CH$^+$ 
feature at 4232.4 \AA{} has the greatest impact on the resulting 
$^{12}$CH$^+$/$^{13}$CH$^+$ ratio, but this variation is less than 10\% of the 
$^{13}$CH$^+$ line strength, which is well within the 20\% observational errors 
associated with the detection of this weak feature.

\begin{figure}[!t]
\centering
\includegraphics[width=0.45\textwidth]{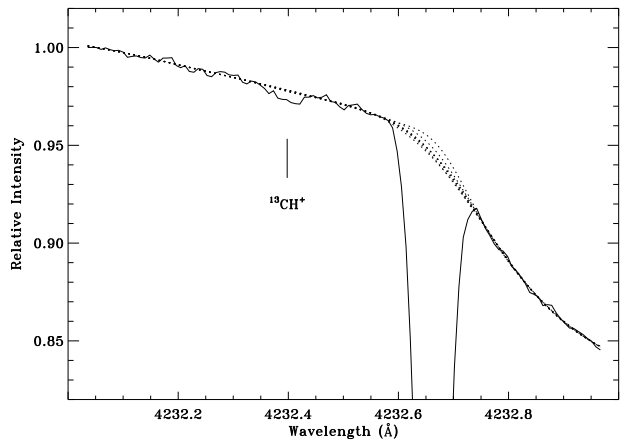}
\caption[Possible continuum fits to the region surrounding the CH$^+$ lines 
toward 20~Tau.]{Possible continuum fits (\emph{dotted lines}) to the region 
surrounding the CH$^+$ lines toward 20~Tau. The strong interstellar 
$^{12}$CH$^+$ line is superimposed on the blue wing of a photospheric 
Fe~{\small II} line, complicating the placement of the continuum. The position 
of the $^{13}$CH$^+$ line is indicated by a tick mark.}
\end{figure}

\subsection{VLT/UVES Archival Spectra}
To supplement our high signal-to-noise McDonald observations of bright targets, 
optical spectra of seven fainter stars were obtained from the European Southern 
Observatory (ESO) Science Archive Facility. These spectra were acquired by the 
Ultraviolet and Visual Echelle Spectrograph (UVES) of the Very Large Telescope 
(VLT) at Cerro Paranal, Chile primarily during two observing programs designed 
to measure $^{12}$CH$^+$/$^{13}$CH$^+$ ratios (see Casassus et al. 2005; Stahl 
et al. 2008). The primary UVES data (from programs 071.C-0367 and 076.C-0431; 
S. Casassus, PI) have S/N ratios that are comparable to those of our McDonald 
spectra. These observations employed the central wavelength setting at 4370 
\AA, which allows continuous spectral coverage from 3750$-$5000 \AA{}, a range 
that includes the CN (0,~0), CH$^+$ (0,~0), and CH$^+$ (1,~0) lines. Other 
VLT/UVES datasets were obtained (from program 065.I-0526; E. Roueff, PI) that 
were acquired using the wavelength setting at 3460~\AA. This setting covers the 
region from 3050$-$3870 \AA{} and permits detection of the CN (1,~0) lines near 
3579 \AA. The observations at 3460~\AA{} typically have lower S/N than those at 
4370 \AA. For one sight line (HD~210121), observations at 3460 \AA{} and 4370 
\AA{} were provided by program 065.I-0526 and additional data acquired with the 
wavelength setting at 3900 \AA{} were obtained from program 071.C-0513 (M. 
Andr\'e, PI).

The weaker CN and CH$^+$ (1,~0) lines provide a check on the column densities 
derived from the stronger (0,~0) lines, which is especially important when the 
(0,~0) lines become optically thick. The seven UVES sight lines chosen for this 
study (see Table 1) are known from previous observations (e.g., Gredel et al. 
1991; Palazzi et al. 1992) to have strong lines of interstellar CN and 
determining $^{12}$CN/$^{13}$CN ratios will be our focus. We do not rederive 
$^{12}$CH$^+$/$^{13}$CH$^+$ ratios for the UVES sight lines as there are 
significant uncertainties associated with such determinations from these data 
(see \S{}~3). While the principal reason for incorporating the UVES spectra 
into our analysis was to expand our sample to include more heavily-reddened 
sight lines, the VLT observing programs also provided us with additional data 
on some of our McDonald targets. Both 23~Tau and $\zeta$~Oph were observed at 
4370 \AA{} under programs 076.C-0431 and 071.C-0367, respectively, and 
observations of $\rho$~Oph~A and $\zeta$~Oph at 3460 \AA{} are available from 
program 065.I-0526. These supplementary data are useful for making comparisons 
with our McDonald results.

All VLT/UVES datasets were reduced with the UVES pipeline software in optimal 
extraction mode unless the observations employed an image slicer in which case 
the average extraction method had to be adopted. Master bias and master flat 
frames were created for each night for which we obtained science observations. 
Th-Ar lamp spectra acquired contemporaneously with the science data were used 
for wavelength calibration. After extraction, all echelle orders from a given 
exposure were merged and the resulting spectra were shifted to the LSR frame of 
reference. Finally, all individual exposures of the same target obtained with 
the same instrumental setup were coadded. The spectral resolution of the UVES 
data varies somewhat for the different observing programs due to changes in 
slit width and CCD binning. The primary data at 4370 \AA{} have a velocity 
resolution of $\Delta$$v=3.5$ km~s$^{-1}$, corresponding to a resolving power 
of $R=85,000$. The observations at 3460 \AA{} employed either 1$\times$1 or 
1$\times$2 binning and have resolutions of 3.8~km~s$^{-1}$ ($R=79,000$) or 
4.2~km~s$^{-1}$ ($R=71,000$), respectively. For HD~210121, the observations at 
4370~\AA{} were combined with those at 3900 \AA{} in the overlapping region, 
and the resulting spectrum has a resolution corresponding to $R=80,000$.

The coadded UVES spectra in the vicinity of CN were normalized to the continuum 
in much the same way as were the McDonald data except that a larger spectral 
window was used (typically 5 \AA) that included all of the lines of the CN 
(0,~0) or (1,~0) band. The $R$(0) and $R$(1) lines in these data are separated 
by very little continuum due to the lower resolution. The larger spectral 
windows, in addition to a non-negligible amount of residual fringing in the 
extracted spectra, meant that higher-order polynomials had to be applied during 
the normalization process. Moreover, the $^{12}$CN and $^{13}$CN $R$(0) lines of 
the (0,~0) band, which are separated by approximately 13.5 km s$^{-1}$, are 
partially blended in the UVES spectra, again a result of the coarse resolution 
but also due to the strength of CN absorption along these lines of sight. 
Absorption from $^{13}$CN is thus found on the redward wing of the $^{12}$CN 
line (as shown in Figure 2), adding some uncertainty to the placement of the 
continuum in this region. The situation is not as problematic as in the case of 
CH$^+$ from these data, however, because the CN lines are intrinsically narrow 
and exhibit mostly single velocity components.

\begin{figure}[!t]
\centering
\includegraphics[width=0.45\textwidth]{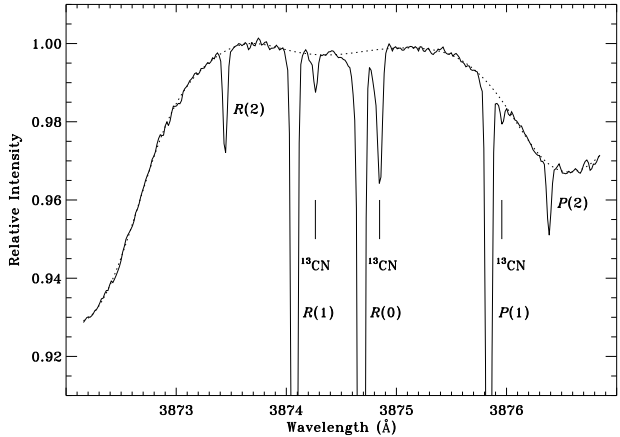}
\caption[Coadded UVES spectrum of CN toward HD~169454 showing the adopted 
continuum fit.]{Coadded UVES spectrum of CN $B$$-$$X$ (0,~0) toward HD~169454 
showing the adopted continuum fit (\emph{dotted line}). Each line is labeled 
and the positions of the $R$(0), $R$(1), and $P$(1) lines of $^{13}$CN are 
indicated by tick marks. This is the first reported detection of the $^{13}$CN 
$P$(1) line in the interstellar medium. Note also that absorption from 
$^{13}$CN $R$(0) is partially blended with the much stronger $^{12}$CN $R$(0) 
line.}
\end{figure}

\subsection{Stellar Sample and Observational Results}
The total exposure times and signal-to-noise ratios per resolution element 
achieved for each of the McDonald and UVES targets are shown in Table 1 along 
with the relevant stellar data, consisting of the spectral type, $B$ magnitude, 
$E$($B-V$) color excess, Galactic $l$ and $b$ coordinates, and distance. We 
list the \emph{Hipparcos} parallax distance (Perryman et al. 1997) if the 
measurement is significant at the 4-$\sigma$ level or greater. Otherwise, the 
distance we derive from the method of spectroscopic parallax is given. The 
spectral types in Table 1 are from the same references that provided values of 
$E$($B-V$), although, for HD~73882, HD~154368, and HD~170740, the spectral 
types were actually listed in a previous paper by the same group (Rachford et 
al. 2002). From these references, we also obtained the $V$ magnitude and total 
visual extinction ($A_V$) for each star without a precise \emph{Hipparcos} 
measurement so that spectroscopic distances could be derived using a 
self-consistent set of photometric data. As the distances in Table 1 indicate, 
our selection of interstellar sight lines samples almost exclusively the local 
ISM. Only four stars in the sample (HD~73882, HD~152236, HD~154368, and 
HD~169454) lie farther than 500 pc from the Sun. Accordingly, these stars also 
exhibit the highest degrees of interstellar reddening, with the largest value 
of $E$($B-V$) found in the direction of HD~169454. Even along the more extended 
sight lines, however, the clouds responsible for CN and CH$^+$ absorption are 
likely to be nearby. This can be deduced from the velocities of the molecular 
components, which correspond to the dominant components in species like 
Ca~{\small II}. The more distant stars have rather more complicated profiles of 
Ca~{\small II} $\lambda$3933, but the dominant components are always found near 
$v_{\mathrm{LSR}}$~=~0~km~s$^{-1}$, implying that the absorption originates in 
local gas.

The S/N ratios in Table 1 result in uncertainties in equivalent width 
($W_{\lambda}$) of 0.06 m\AA, on average, for the CN and CH$^+$ lines in our 
McDonald spectra. For the primary UVES data, the average uncertainty in 
$W_{\lambda}$(CN) is 0.05 m\AA. Despite the small uncertainties, we do not 
detect absorption from $^{13}$CH$^+$ toward $\zeta$~Per or 20~Aql, which is not 
unexpected considering the weak $^{12}$CH$^+$ lines in these directions. The 
3-$\sigma$ upper limits on the equivalent widths of $^{13}$CH$^+$ are 
$\lesssim0.18$ m\AA{} for $\zeta$~Per and $\lesssim0.24$ m\AA{} for 20~Aql. We 
also do not detect any CN absorption from either isotopologue in the directions 
of 20~Tau and 23~Tau, but obtain upper limits on $W_{\lambda}$($^{12}$CN) of 
$\lesssim0.25$ m\AA{} and $\lesssim0.13$ m\AA{} for the two sight lines, 
respectively. The equivalent widths for detected lines are given in Table 2 for 
CH$^+$ $\lambda$4232 and Table 3 for CN $\lambda$3874. We list the fitted 
equivalent widths that result from our profile synthesis analysis (see \S{} 3). 
Uncertainties in $W_{\lambda}$ were calculated as the product of the rms 
variations in the continuum and the full width at half maximum (FWHM) in the 
line\footnote{While we do not formally include continuum placement errors when 
calculating uncertainties in $W_{\lambda}$, the uncertainties derived using our 
method are more conservative than those based on statistical errors (adopting 
the formula given in Jenkins et al. 1973) and our estimates for continuum 
placement errors added in quadrature. As an example, the statistical error in 
the equivalent width of the $^{13}$CH$^+$ line toward 20~Tau is 
$\sigma_{W,\mathrm{rms}}=(\Delta\lambda)(N_{\mathrm{p}}^{1/2})(\mathrm{rms})=0.0496$
m\AA, where $\Delta\lambda=0.00802$~\AA{} is the wavelength dispersion, 
$N_{\mathrm{p}}=21$ is the number of pixels sampling the line profile, and 
rms~=~0.00135 is the root mean square of the noise in the continuum. The 
standard deviation in the measured value of $W_{\lambda}$($^{13}$CH$^+$) for a 
range of possible continuum fits (see Figure~1) is 
$\sigma_{W,\mathrm{cont}}=0.0229$~m\AA. Thus, the total error would be 
$\sigma_W=(\sigma_{W,\mathrm{rms}}^2+\sigma_{W,\mathrm{cont}}^2)^{1/2}=0.0546$~m\AA, 
while our method gives $\sigma_W=(\mathrm{FWHM})(\mathrm{rms})=0.0730$~m\AA.}. 
Note the quite strong detections of $^{13}$CN toward the UVES 
targets. While not listed in Table 3, the weak $R$(2) and $P$(2) lines of 
$^{12}$CN are detected toward most of the stars in our sample with CN 
absorption. In our McDonald spectra, however, these features are only marginal. 
For $\zeta$~Oph and HD~152236, the lower signal-to-noise ratios near 3874 
\AA{}, compared to other stars observed with the same instrument, prevent any 
detection of $R$(2) or $P$(2). Remarkably, four of the UVES targets exhibit 
fairly significant absorption from $^{13}$CN $R$(1), and in the spectrum of 
HD~169454, the $^{13}$CN $P$(1) line is detected with a significance of 
8~$\sigma$ (see Figure 2). Also shown in the tables are previous results from 
the literature on CH$^+$ and CN toward our targets, where we have included only 
those studies that give equivalent widths for the weaker isotopologue or for 
the $R$(1) and $P$(1) lines in the case of CN. Studies that examined only the 
strongest line in CH$^+$ or CN are omitted.

\begin{figure}[!t]
\centering
\includegraphics[width=0.45\textwidth]{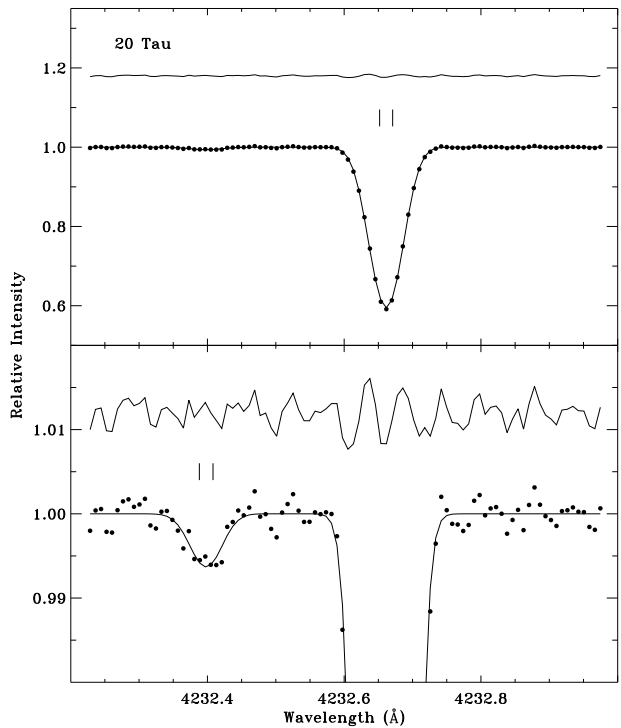}
\caption[Profile synthesis fit to the CH$^+$ lines toward 20~Tau.]{Simultaneous 
profile synthesis fit to the $^{12}$CH$^+$ and $^{13}$CH$^+$ $R$(0) lines toward 
20~Tau scaled to show the stronger (\emph{upper panel}) and weaker (\emph{lower 
panel}) isotopologue. The synthetic profile is shown as a solid line passing 
through data points that represent the observed spectrum. Residuals are plotted 
above the fit. Individual velocity components are indicated by tick marks.}
\end{figure}

\section{PROFILE SYNTHESIS}

The CN and CH$^+$ absorption profiles were synthesized with the rms-minimizing 
code ISMOD (Y. Sheffer, unpublished) that treats the velocities ($v$), Doppler 
$b$-values, and column densities ($N$) of the absorption components as free 
parameters. ISMOD assumes a Voigt profile function for each absorption 
component and convolves the intrinsic profile with an instrumental profile, 
represented by a Gaussian with a width determined by the resolving power of the 
instrument (see Black \& van Dishoeck 1988). The $R$(0) lines of the (0,~0) 
bands in CN and CH$^+$ were used to derive the $N$($^{12}$CN)/$N$($^{13}$CN) and 
$N$($^{12}$CH$^+$)/$N$($^{13}$CH$^+$) ratios. Both isotopologues of CN and of 
CH$^+$ were fitted simultaneously by requiring the component structure of the 
two profiles to be identical. Thus, the 12-to-13 column density ratio and the 
isotope shift between the two lines were additional free parameters.

\begin{figure}[!t]
\centering
\includegraphics[width=0.45\textwidth]{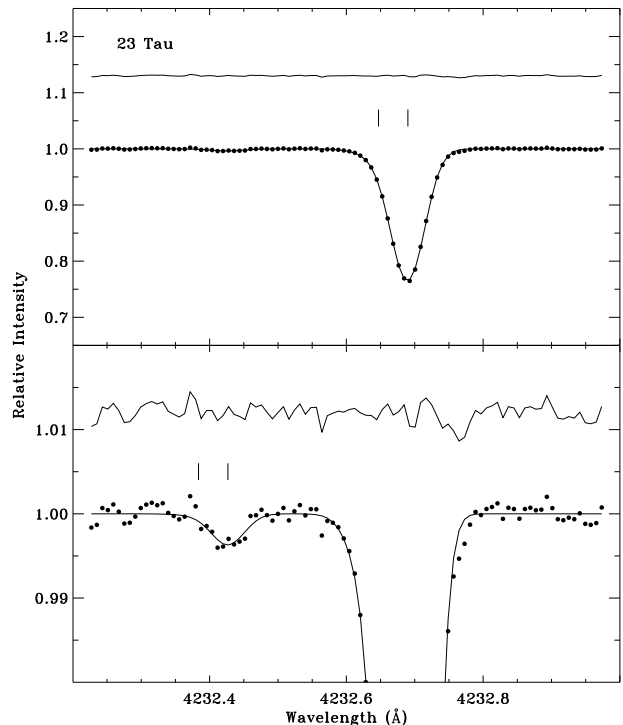}
\caption[Profile synthesis fit to the CH$^+$ lines toward 23~Tau.]{Same as 
Figure 3 except for the $^{12}$CH$^+$ and $^{13}$CH$^+$ $R$(0) lines toward 
23~Tau.}
\end{figure}

\begin{figure}[!t]
\centering
\includegraphics[width=0.45\textwidth]{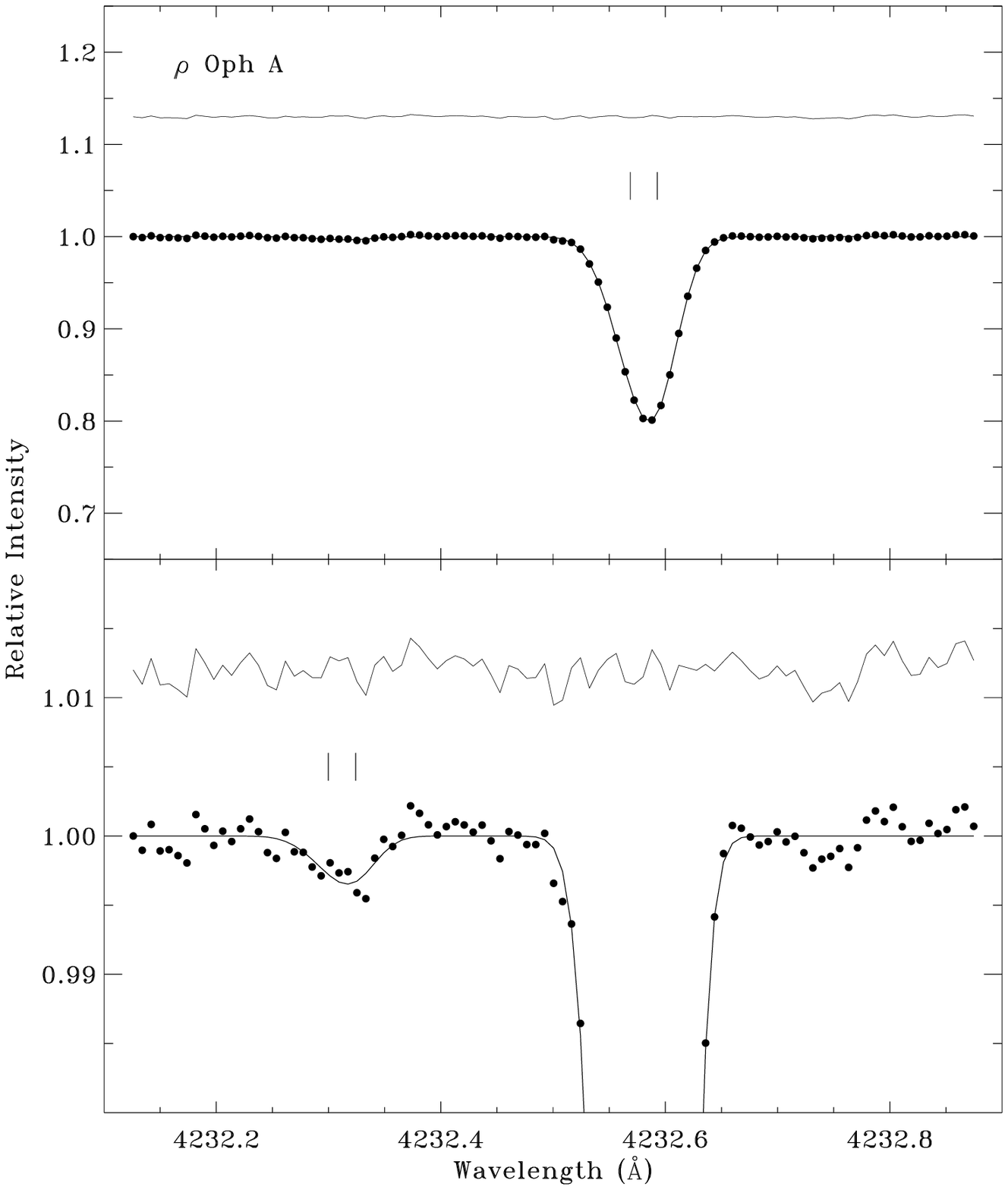}
\caption[Profile synthesis fit to the CH$^+$ lines toward $\rho$~Oph~A.]{Same 
as Figure 3 except for the $^{12}$CH$^+$ and $^{13}$CH$^+$ $R$(0) lines toward 
$\rho$~Oph~A.}
\end{figure}

\begin{figure}[!t]
\centering
\includegraphics[width=0.45\textwidth]{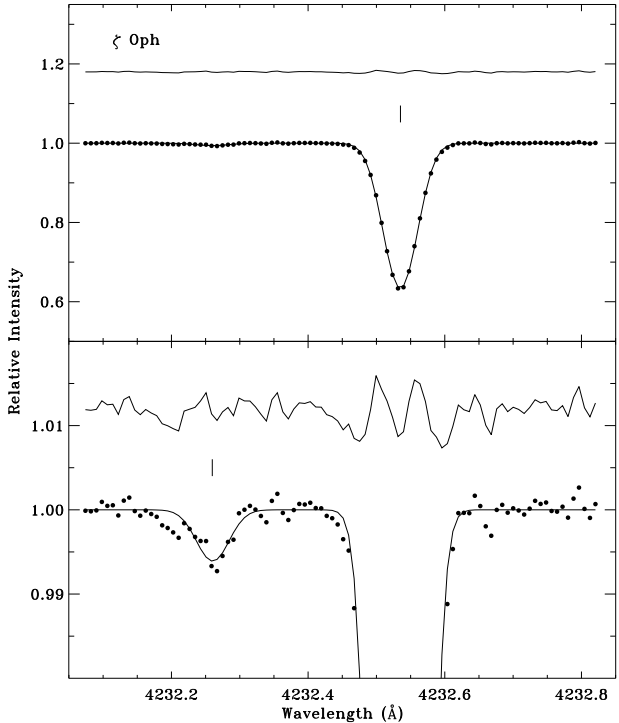}
\caption[Profile synthesis fit to the CH$^+$ lines toward $\zeta$~Oph.]{Same as 
Figure 3 except for the $^{12}$CH$^+$ and $^{13}$CH$^+$ $R$(0) lines toward 
$\zeta$~Oph.}
\end{figure}

Figures 3 through 10 present the CH$^+$ and CN $R$(0) fits for the data 
acquired at McDonald Observatory. When fitting the CH$^+$ lines toward 20~Tau 
and 23~Tau, we used our previous results on the component structure in these 
directions (Ritchey et al. 2006) as the initial input to the profile synthesis 
routine. Likewise, the initial values for the component structure of CH$^+$ 
toward $\rho$~Oph~A were obtained from Pan et al. (2004). The input values for 
the CN $R$(0) line toward $\zeta$~Oph were taken from the ultra-high resolution 
($R=600,000$) study by Lambert et al. (1990), who modeled CN absorption along 
this sight line with two Gaussian components separated by 1.2 km s$^{-1}$. 
Because our study has coarser resolution and the component parameters in this 
case were already precisely determined, the velocity difference between the two 
components, their relative strengths, and their $b$-values were held fixed in 
our analysis. Lambert et al. (1990) also examined CH$^+$ toward $\zeta$~Oph, 
but modeled the line with a single Gaussian component. Absorption profiles with 
single components do not require precise input parameters in ISMOD. Thus, for 
the CH$^+$ line toward $\zeta$~Oph, as well as for the CN $R$(0) lines toward 
the other McDonald targets, input parameters were determined by direct 
examination of the spectra within IRAF. The velocities and column densities of 
the interstellar absorption components in our profile synthesis fits were 
computed using the wavelengths and oscillator strengths ($f$-values) presented 
in Table 4.

\begin{figure}[!t]
\centering
\includegraphics[width=0.45\textwidth]{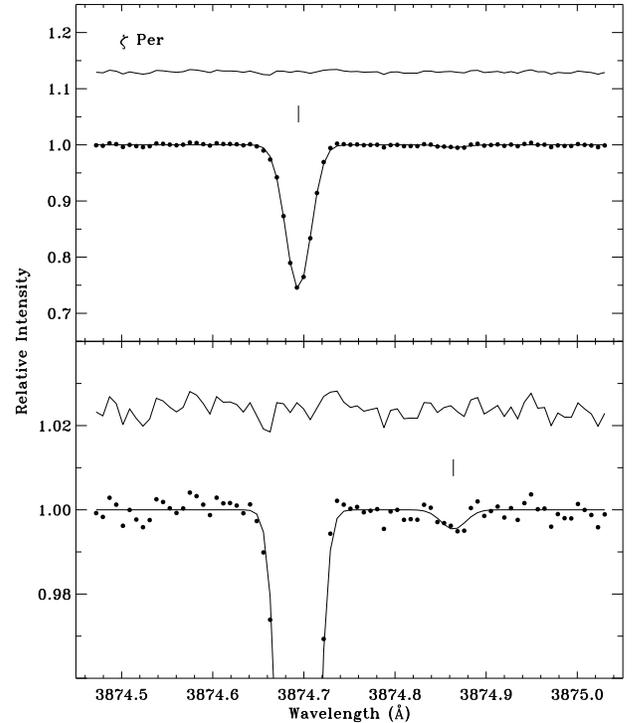}
\caption[Profile synthesis fit to the CN lines toward $\zeta$~Per.]
{Simultaneous profile synthesis fit to the $^{12}$CN and $^{13}$CN $R$(0) lines 
toward $\zeta$~Per scaled to show the stronger (\emph{upper panel}) and weaker 
(\emph{lower panel}) isotopologue. Plotting symbols are the same as in 
Figure~3.}
\end{figure}

\begin{figure}[!t]
\centering
\includegraphics[width=0.45\textwidth]{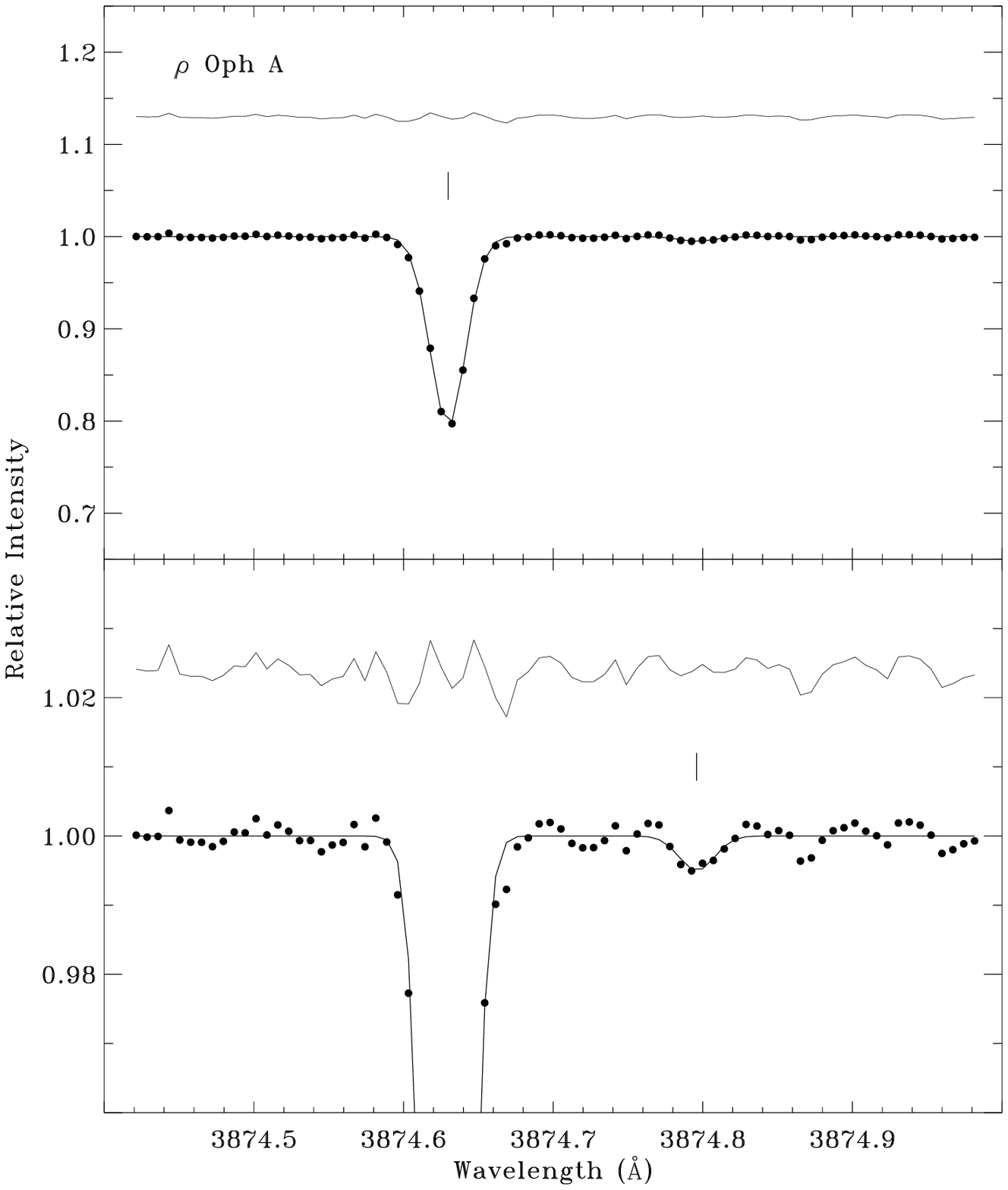}
\caption[Profile synthesis fit to the CN lines toward $\rho$~Oph~A.]{Same as 
Figure 7 except for the $^{12}$CN and $^{13}$CN $R$(0) lines toward 
$\rho$~Oph~A.}
\end{figure}

\begin{figure}[!t]
\centering
\includegraphics[width=0.45\textwidth]{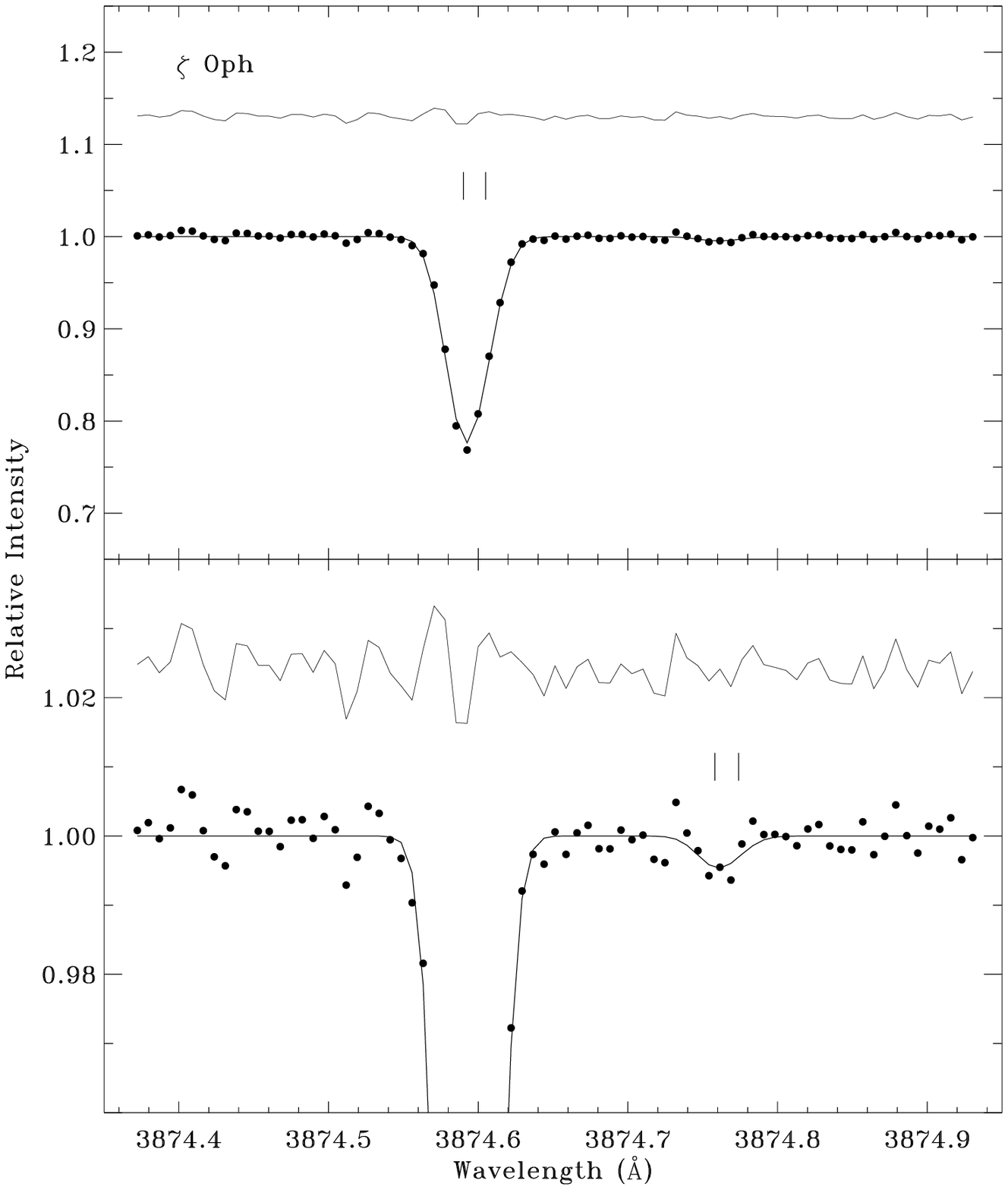}
\caption[Profile synthesis fit to the CN lines toward $\zeta$~Oph.]{Same as 
Figure 7 except for the $^{12}$CN and $^{13}$CN $R$(0) lines toward 
$\zeta$~Oph.}
\end{figure}

\begin{figure}[!t]
\centering
\includegraphics[width=0.45\textwidth]{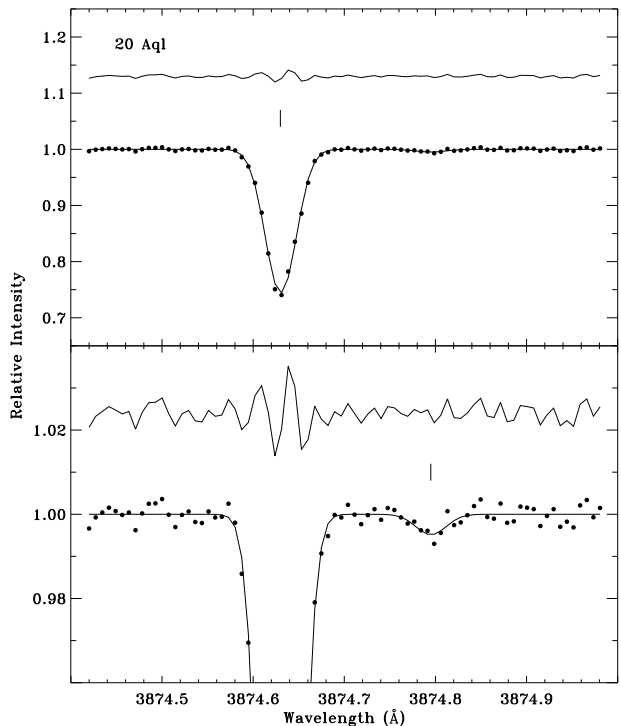}
\caption[Profile synthesis fit to the CN lines toward 20~Aql.]{Same as Figure 7 
except for the $^{12}$CN and $^{13}$CN $R$(0) lines toward 20~Aql.}
\end{figure}

Among the optical tracers of diffuse molecular gas, the CN molecule typically 
probes the densest portion of a diffuse cloud (Pan et al. 2005). As a 
consequence, CN line widths are intrinsically small, corresponding to 
$b$-values that are less than 1.0 km s$^{-1}$ in most cases. With such narrow 
lines, the strongest transitions in CN can become optically thick even at 
moderate column densities. While this is not a major concern for our McDonald 
sight lines, all of which have optical depths ($\tau$) at line center that are 
less than 1.0, it is an important consideration for the UVES observations, 
where the average value of $\tau$ for the (0,~0) $R$(0) line is 4.45. When 
large optical depths are present, a small change in $b$(CN) will have a 
dramatic impact on the $^{12}$CN column density obtained from the strong $R$(0) 
line, and the resulting $^{12}$CN/$^{13}$CN ratio will be similarly affected. 
Moreover, because the UVES spectra are sampled at moderate resolution 
($\sim$3.5~km~s$^{-1}$), the CN lines are dominated by instrumental rather than 
intrinsic broadening, making direct evaluation of the $b$-values more 
difficult. To minimize these uncertainties, $b$-values for optically thick 
lines were derived using the well-known doublet ratio method, which has often 
been employed in this context (e.g., Meyer \& Jura 1985; Gredel et al. 1991; 
Roth \& Meyer 1995). The method is predicated on the fact that two lines 
arising from the same state must yield identical column densities for some 
value of $b$. In the case of CN, one can use either the (0,~0) $R$(1) and 
$P$(1) lines or the (0,~0) and (1,~0) $R$(0) lines. Having observations of both 
pairs is advantageous as it allows independent checks on consistency.

In the optically thin limit, the equivalent width of an absorption line is 
proportional to the column density such that $W_{\lambda} \propto \lambda^2 f N$ 
(see Spitzer 1978). Thus, for a pair of optically thin lines that must yield 
identical values of $N$, the equivalent width ratio is given by 
$W_{\lambda,1}/W_{\lambda,2} = (\lambda_{1}/\lambda_{2})^2 f_{1}/f_{2}$. 
Specifically, from the molecular line parameters in Table 4, the (0,~0) 
$R$(1)/$P$(1) equivalent width ratio for optically thin absorption is equal to 
2.00, while the (0,~0)/(1,~0) $R$(0) ratio is equal to 13.35. The measured 
ratio will be lower if the optical depth of the stronger line is significant. 
To derive $b$-values using this approach, the CN equivalent widths were 
determined by fitting simple Gaussians to the observed profiles within IRAF. 
Column densities based on the measured equivalent widths for a given pair of 
lines were then interpolated from a set of curves of growth calculated on the 
interval $0.30 \leq b \leq 0.99$ km s$^{-1}$ with a step size of 0.01 
km~s$^{-1}$. The result of this procedure was a pair of curves that give $N$ as 
a function of $b$ for the two transitions, corresponding to the measured values 
of $W_{\lambda}$. By plotting these curves and finding the intersection in the 
$N$-$b$ plane, one finds the unique $b$-value that yields identical column 
densities for the pair of lines (see Figure~11).

\begin{figure}[!t]
\centering
\includegraphics[width=0.45\textwidth]{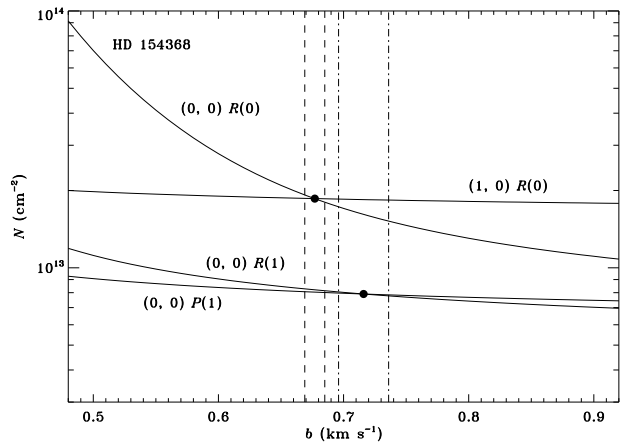}
\caption[Column density as a function of $b$-value for CN transitions toward 
HD~154368.]{Column density as a function of $b$-value for the four CN 
transitions used to derive $b$ in the main component toward HD~154368 via the 
doublet ratio method (see text). The intersection of each pair of curves, which 
gives the $b$-value corresponding to the measured doublet ratio, is marked by a 
solid point. The 1-$\sigma$ errors on $b$ thus derived are shown as dashed 
lines for the (0,~0)/(1,~0) $R$(0) pair and dot-dashed lines for the (0,~0) 
$R$(1)/$P$(1) pair.}
\end{figure}

Table 5 gives the observed CN (0,~0) $R$(1)/$P$(1) equivalent width ratios, and 
(0,~0)/(1,~0) $R$(0) ratios if possible, for all of the stars in our sample 
where CN is detected. For the main CN component toward HD~152236, both ratios 
are consistent with the optically thin limit. The $R$(1)/$P$(1) ratio of 
$1.91\pm0.10$ nominally gives a $b$-value of 0.39 km s$^{-1}$, but with an 
upper bound at infinity. An infinite $b$-value in this context means that the 
lines fall on the linear portion of the curve of growth and cannot be used to 
constrain the Doppler parameter. This is evidently the case for the CN lines 
toward our McDonald targets, which yield no definitive information on $b$. It 
is worth noting, however, that the (0,~0)/(1,~0) $R$(0) equivalent width ratios 
toward $\rho$~Oph~A and $\zeta$~Oph would result in $b$-values of 0.60 
km~s$^{-1}$ and 0.88 km s$^{-1}$, respectively. Even if the uncertainties give 
infinite upper bounds, the nominal results are similar to the results of many 
earlier investigations. For instance, Palazzi et al. (1992) find 
$b=0.84\pm0.22$ km~s$^{-1}$ in the direction of $\rho$~Oph~A, while Pan et al. 
(2004) list a value of 0.7 km s$^{-1}$. For $\zeta$~Oph, Crane et al. (1986) 
give $b=0.88\pm0.02$ km s$^{-1}$, and Roth \& Meyer (1995) find $b=0.89\pm0.19$ 
km~s$^{-1}$. The CN absorption profiles toward HD~161056 arise from two 
components of approximately equal strength, separated by 5.9 km s$^{-1}$. 
Interestingly, the $R$(1)/$P$(1) ratio for the component at $-$3.1 km s$^{-1}$ 
indicates that this component has a slightly larger optical depth than the 
component at +2.8 km s$^{-1}$, although the (0,~0)/(1,~0) $R$(0) ratios in both 
cases exceed the optically thin limit.

The UVES sight lines in Table 5 other than HD~152236 and HD~161056 exhibit 
fairly significant optical depths in their CN absorption lines. An extreme case 
is presented by the sight line to HD~169454, for which the (0,~0)/(1,~0) $R$(0) 
equivalent width ratio is only $2.83\pm0.05$. Where measurements on both pairs 
of CN transitions are available, the derived $b$-values agree remarkably well. 
Note that the ratio between the (0,~0) $R$(0) and (1,~0) $R$(0) equivalent 
widths provides the most stringent constraints on $b$ because of the larger 
difference in intrinsic line strength for this pair, compared to $R$(1) and 
$P$(1). The weighted mean of the two $b$-values in the case of HD~154368 is 
$b=0.68\pm0.01$ km s$^{-1}$, while for HD~169454, HD~170740, and HD~210121 the 
weighted means are $b=0.50\pm0.01$ km~s$^{-1}$, $b=0.46\pm0.04$ km~s$^{-1}$, and 
$b=0.80\pm0.02$ km~s$^{-1}$, respectively. Palazzi et al. (1990) obtained a 
$b$-value of $0.65\pm0.01$ km s$^{-1}$ for HD~154368, and the study by Roth \& 
Meyer (1995) resulted in a $b$-value of $0.68\pm0.08$ km s$^{-1}$ in this 
direction. Additionally, Palazzi et al. (1992) find $b=0.64\pm0.17$ km s$^{-1}$ 
for HD~169454, and $b=0.84\pm0.20$ km s$^{-1}$ for HD~170740. Sheffer et al. 
(2008) give the $b$-value toward HD~210121 as 0.9 km s$^{-1}$. These various 
determinations show good agreement, except in the case of HD~170740. Yet, even 
for this sight line, the discrepancy is not that severe. The Palazzi et al. 
(1992) $b$-value actually agrees with our more precise determination within 2 
$\sigma$ and our result is particularly secure because it was derived from 
measurements of four different CN transitions across two distinct bands, which 
were observed independently. Indeed, as a result of the excellent agreement 
between $b$-values determined from the two equivalent width ratios, the 
weighted means for HD~154368, HD~169454, HD~170740, and HD~210121 were used as 
fixed input in the profile fitting analysis. For HD~73882, HD~152236, and 
HD~161056, the $b$-values obtained from the $R$(1)/$P$(1) ratios were used as 
the initial input to ISMOD but were allowed to vary.

The component parameters resulting from our profile synthesis fits to the 
CH$^+$ and CN $R$(0) lines of the (0,~0) bands are presented in Table 6. In 
cases where isotopologic ratios were determined (see Table 7), the quoted 
velocities and $b$-values apply to both the $^{12}$C and $^{13}$C-bearing 
isotopologues since the two profiles were required to have identical component 
structure. Uncertainties in $N$ for a given component were derived from the 
relative uncertainties in $W_{\lambda}$. For 20~Tau and 23~Tau, the new results 
are consistent with our previous analysis of CH$^+$ absorption in the Pleiades 
from McDonald spectra with much lower signal to noise (Ritchey et al. 2006). 
The only major difference is that, in the present investigation, we derive a 
10\% higher column density of $^{12}$CH$^+$ toward 20~Tau, even though the 
equivalent widths agree to within 2\%. As profile synthesis was not used 
previously to obtain column densities, the discrepancy underscores the 
importance of carefully considering optical depth effects when dealing with 
strong absorption lines. For 23~Tau, both the equivalent widths and column 
densities of $^{12}$CH$^+$ from this and the previous study are consistent at 
the 1\% level. Our new results on CH$^+$ toward $\rho$~Oph~A are also broadly 
consistent with the analysis of Pan et al. (2004), who give a total 
$^{12}$CH$^+$ column density that differs from our value by less than 1\%. In 
the case of CN toward $\rho$~Oph~A, our profile fit results in a $b$-value of 
0.4 km s$^{-1}$, somewhat lower than the 0.7 km s$^{-1}$ given by Pan et al. 
(2004). Since the equivalent width measurements are essentially identical, the 
difference in $b$ amounts to a 20\% increase in $N$($^{12}$CN) in the present 
analysis.

Table 7 lists the $^{12}$CN/$^{13}$CN and $^{12}$CH$^+$/$^{13}$CH$^+$ column 
density ratios for our sample along with the isotope shifts and total molecular 
column densities derived through profile synthesis of the (0,~0) bands in CN 
and CH$^+$. The uncertainties in $^{12}$CN/$^{13}$CN and 
$^{12}$CH$^+$/$^{13}$CH$^+$ were obtained by propagating the errors associated 
with the column densities of the two isotopologues, although the errors in $N$ 
for the weaker isotopologue essentially determine the outcome. Also included in 
Table 7 are column densities and 12-to-13 ratios for CO from the literature, 
where such measurements exist. In Table 8, we present a comparison between the 
column densities that we derive from the (0,~0) $R$(0) and (1,~0) $R$(0) lines 
of CH$^+$ and CN. The good agreement indicates that we have adequately 
accounted for optical depth effects in these spectra, giving us confidence in 
our isotopologic results.

As mentioned in \S{}~2.2, we are not confident that $^{12}$CH$^+$/$^{13}$CH$^+$ 
ratios can be reliably obtained for the UVES sight lines. A number of 
contributing factors lead us to this assessment. There is a non-negligible 
amount of fringing in the reduced UVES spectra, making an unambiguous placement 
of the continuum impossible. The problem is compounded by the broad and 
moderately complex CH$^+$ absorption profiles, which often consist of three or 
more components (see Table 6). Particularly worrisome are weak $^{12}$CH$^+$ 
components that lie to the blue of the main component (for instance, toward 
HD~152236, HD~154368, and HD~169454). In the (0,~0) line at 4232 \AA{}, these 
components are likely to be blended with the strongest feature in 
$^{13}$CH$^+$, considering that the isotope shift is $-19$~km~s$^{-1}$. In 
principle, one could constrain the $^{12}$CH$^+$ absorption profile using the 
(1,~0) transition at 3957 \AA{} since the $^{13}$CH$^+$ line corresponding to 
this transition is shifted by 33 km s$^{-1}$ to the red of $^{12}$CH$^+$, rather 
than to the blue. In practice, however, components that are already weak at 
4232~\AA{} may not even be detected in the weaker transition at 3957 \AA, 
limiting the usefulness of this approach.

\begin{figure*}[!t]
\centering
\includegraphics[angle=90,width=0.95\textwidth]{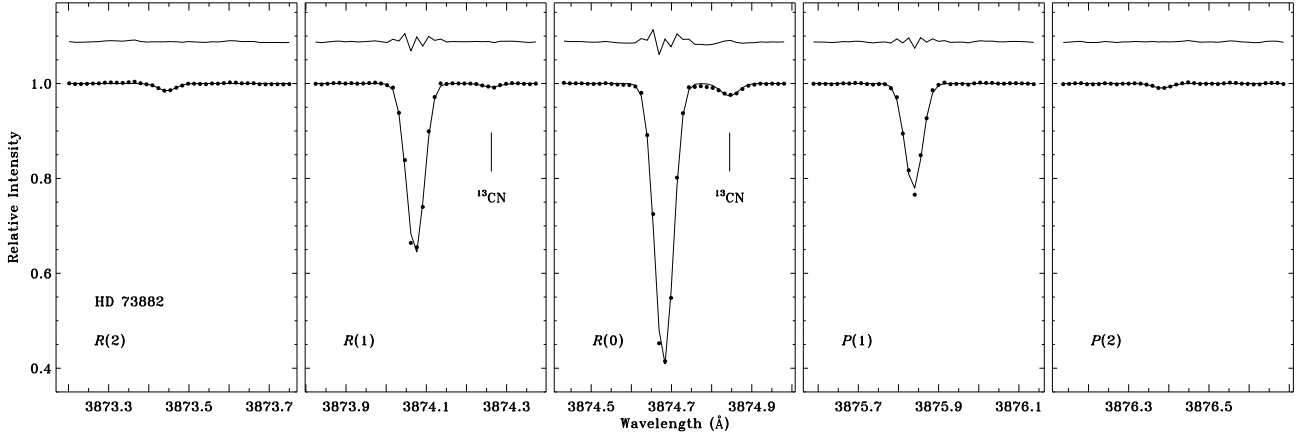}
\caption[Profile synthesis fits to the CN lines toward HD~73882.]{Profile 
synthesis fits to the $B$$-$$X$ (0,~0) band of CN toward HD~73882. See Figure 3 
for a description of the plotting symbols. The same range in velocity is shown 
for each panel. The positions of the $R$(0) and $R$(1) lines of $^{13}$CN are 
indicated by tick marks. When both isotopologues are present, the two profiles 
are synthesized simultaneously.}
\end{figure*}

\begin{figure*}[!t]
\centering
\includegraphics[angle=90,width=0.95\textwidth]{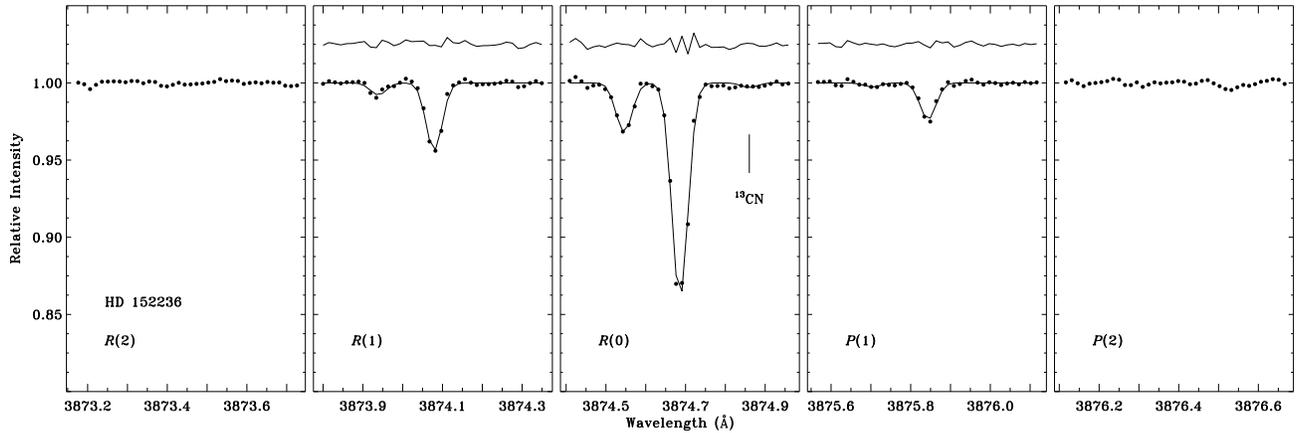}
\caption[Profile synthesis fits to the CN lines toward HD~152236.]{Same as 
Figure 12 except for the CN $B$$-$$X$ (0,~0) band toward HD~152236. Two 
line-of-sight components are clearly identified in the $R$(0) and $R$(1) lines 
of $^{12}$CN and are also detected in the $P$(1) line. The position of the 
$^{13}$CN $R$(0) line corresponding to the stronger of the two components is 
indicated by a tick mark. Since the $R$(2) and $P$(2) lines are below the 
detection limit, no fits were attempted for these features.}
\end{figure*}

Ultimately, we did synthesize the CH$^+$ profiles toward the UVES targets in an 
attempt to derive $^{12}$CH$^+$/$^{13}$CH$^+$ ratios, but the results bore out 
our initial concerns. No $^{13}$CH$^+$ absorption is detected toward HD~152236 
nor toward HD~210121 due to the low column densities of the individual 
$^{12}$CH$^+$ components in these directions. For HD~154368, HD~161056, and 
HD~169454, the $^{12}$CH$^+$ absorption profiles at 4232~\AA{} are almost 
certainly blended with $^{13}$CH$^+$, making it difficult to obtain any 
meaningful results on the $^{12}$CH$^+$/$^{13}$CH$^+$ ratio. For sight lines 
with simpler profiles, the results are more encouraging, but the uncertainties 
are still quite large. The $^{12}$CH$^+$ and $^{13}$CH$^+$ features in the 
spectrum of HD~170740 do not appear to be blended, but $^{13}$CH$^+$ is only 
detected at the 1.4-$\sigma$ level. The sight line to HD~73882 permits a 
relatively clean detection of both $^{12}$CH$^+$ and $^{13}$CH$^+$ at 
4232~\AA{}, though the continuum level is somewhat uncertain. Nevertheless, we 
find a $^{12}$CH$^+$/$^{13}$CH$^+$ ratio in this direction of $61.7\pm17.3$ for 
a fixed isotope shift of $\Delta v=-18.9$~km~s$^{-1}$.

We also analyzed the available UVES data on CH$^+$ toward 23~Tau and 
$\zeta$~Oph, both of which have simple, well-studied CH$^+$ absorption 
profiles. From the (0,~0) transition at 4232 \AA, we obtained a 
$^{12}$CH$^+$/$^{13}$CH$^+$ ratio of $102.3\pm38.4$ for 23~Tau 
($\Delta v=-19.4$~km~s$^{-1}$) and $88.5\pm11.4$ for $\zeta$~Oph 
($\Delta v=-19.0$~km~s$^{-1}$). However, independent fits to the (1,~0) line at 
3957 \AA{} yielded lower values ($55.5\pm29.4$ for 23~Tau and $52.8\pm9.1$ for 
$\zeta$~Oph). Taking the weighted mean of the two results in each case gives 
$72.8\pm23.3$ for 23~Tau and $66.7\pm7.1$ for $\zeta$~Oph, both of which are 
consistent with the $^{12}$CH$^+$/$^{13}$CH$^+$ ratios we derive from our 
McDonald observations of these stars (see Table 7). Still, the disparity in the 
values obtained from the two transitions highlights the problems encountered 
when attempting to fit the undulating continuum in the UVES spectra examined 
here. Given the uncertainties, our UVES results are not altogether dissimilar 
to those of Casassus et al. (2005) and Stahl et al. (2008), who first analyzed 
these data. Casassus et al. found a $^{12}$CH$^+$/$^{13}$CH$^+$ ratio of 
$78.5\pm3.5$ for the sight line to $\zeta$~Oph, while Stahl et al., reanalyzing 
the $\zeta$~Oph data, obtained a ratio of $80.9\pm3.0$. In the direction of 
23~Tau, Stahl et al. found $^{12}$CH$^+$/$^{13}$CH$^+$ = $96.5\pm16.3$, while 
toward HD~73882, these authors obtained a $^{12}$CH$^+$/$^{13}$CH$^+$ ratio of 
$79.7\pm14.5$.

\begin{figure*}[!t]
\centering
\includegraphics[angle=90,width=0.95\textwidth]{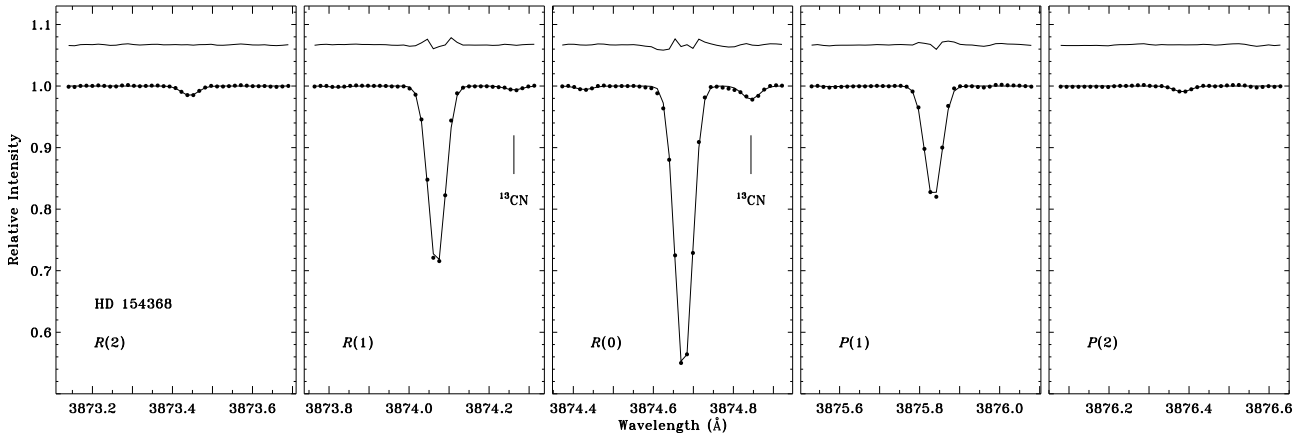}
\caption[Profile synthesis fits to the CN lines toward HD~154368.]{Same as 
Figure 12 except for the CN $B$$-$$X$ (0,~0) band toward HD~154368. The 
positions of the $R$(0) and $R$(1) lines of $^{13}$CN are indicated by tick 
marks. The weak absorption feature blueward of the strong $^{12}$CN $R$(0) line 
is due to a second very weak component along the line of sight, which is 
positively identified in the spectra of CH$^+$ and CH.}
\end{figure*}

\begin{figure*}[!t]
\centering
\includegraphics[angle=90,width=0.95\textwidth]{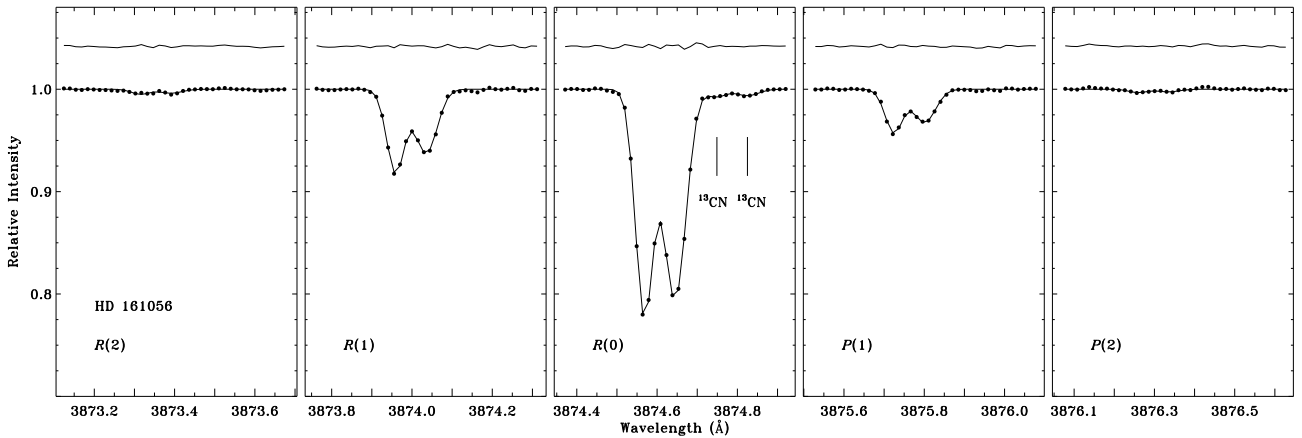}
\caption[Profile synthesis fits to the CN lines toward HD~161056.]{Same as 
Figure 12 except for the CN $B$$-$$X$ (0,~0) band toward HD~161056. Two 
components of approximately equal strength are identified in each of the 
$^{12}$CN lines. The positions of the $^{13}$CN $R$(0) lines corresponding to 
these components are indicated by tick marks.}
\end{figure*}

The weaker lines of the $B$$-$$X$ (0,~0) band in CN were fitted using the 
component structure found for the $R$(0) line. Since the majority of CN 
absorption profiles consist of single velocity components, this restriction 
essentially amounted to keeping the $b$-values fixed. In cases where an 
additional component is observed, the velocity difference between the two 
components, as determined from the $R$(0) line, was also held constant. The 
strengths of the individual components were allowed to vary, however, so that 
excitation temperatures could be derived separately for each cloud along the 
line of sight. Although two lines arising from the same excitation state, such 
as $R$(1) and $P$(1) or $R$(2) and $P$(2), should yield identical column 
densities, each line was synthesized independently in our analysis. The final 
column density for a given $N=1$ or $N=2$ level was then determined by taking 
the weighted mean of the two results. Table 9 presents the CN column densities 
for each rotational level with a detectable transition. For the four UVES sight 
lines where $^{13}$CN $R$(1) is detected, the $R$(1) lines of both 
isotopologues were fitted simultaneously as in the case of $R$(0). The same 
technique was applied to the $P$(1) lines of $^{12}$CN and $^{13}$CN in the 
direction of HD~169454. Because this procedure permits the 12-to-13 ratios in 
different $N$ levels to vary, we can accomodate small variations in excitation 
temperature between $^{12}$CN and $^{13}$CN. Figures 12 through 18 give the 
profile synthesis fits to the CN $B$$-$$X$ (0,~0) band for each of the UVES 
sight lines.

\begin{figure*}[!t]
\centering
\includegraphics[angle=90,width=0.95\textwidth]{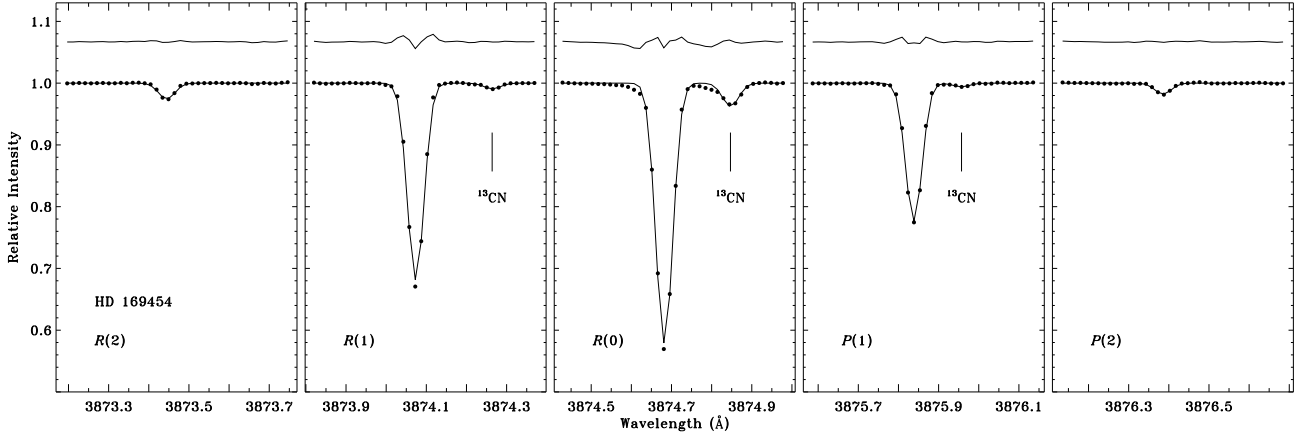}
\caption[Profile synthesis fits to the CN lines toward HD~169454.]{Same as 
Figure 12 except for the CN $B$$-$$X$ (0,~0) band toward HD~169454. The 
positions of the $R$(0), $R$(1), and $P$(1) lines of $^{13}$CN are indicated by 
tick marks.}
\end{figure*}

\begin{figure*}[!t]
\centering
\includegraphics[angle=90,width=0.95\textwidth]{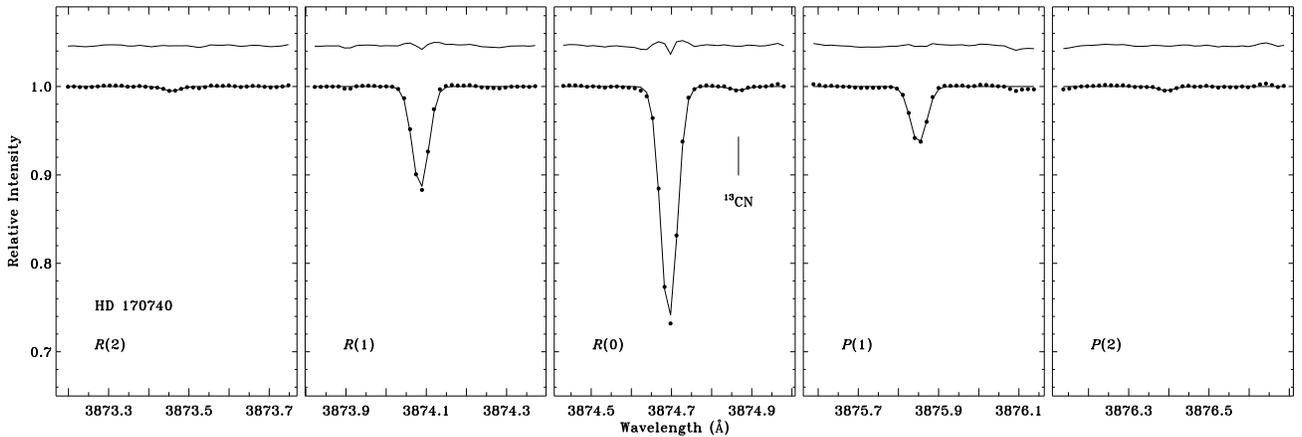}
\caption[Profile synthesis fits to the CN lines toward HD~170740.]{Same as 
Figure 12 except for the CN $B$$-$$X$ (0,~0) band toward HD~170740. The 
position of the $^{13}$CN $R$(0) line is indicated by a tick mark.}
\end{figure*}

Since the isotope shifts in CN and CH$^+$ were left as free parameters during 
profile synthesis, it is instructive to compare our results, from astronomical 
spectra, with those obtained in laboratory experiments. The mean isotope shift 
in our synthesis fits of the CH$^+$ (0,~0) transition at 4232 \AA{} (toward 
McDonald targets) is $\Delta v=-18.9\pm0.4$~km~s$^{-1}$ 
($\delta \lambda=-0.267\pm0.005$ \AA). This result is consistent with the 
experimental value of $-19.3\pm0.3$ km s$^{-1}$ ($-0.272\pm0.004$ \AA), which 
is based on the laboratory measurements of Carrington \& Ramsay (1982) for 
$^{12}$CH$^+$ and Bembenek (1997) for $^{13}$CH$^+$. Similarly, our mean isotope 
shift for the CN (0,~0) $R$(0) line, $\Delta v=+13.1\pm0.4$ km s$^{-1}$ 
($\delta \lambda=+0.169\pm0.005$ \AA), agrees favorably with the laboratory 
value, $\Delta v=+13.6$~km~s$^{-1}$ ($\delta \lambda=+0.176$ \AA; Bakker \& 
Lambert 1998). The four detections of $^{13}$CN $R$(1) yield a mean isotope 
shift for the $R$(1) line of $+14.6\pm0.4$ km s$^{-1}$ ($+0.188\pm0.005$~\AA), 
while, from the $^{12}$CN and $^{13}$CN $P$(1) lines observed toward HD~169454, 
we derive an isotope shift of $+9.2\pm0.4$~km~s$^{-1}$ ($+0.118\pm0.005$ \AA). 
The laboratory results give +14.8~km~s$^{-1}$ (+0.192 \AA) for the $R$(1) line 
blend and +10.0~km~s$^{-1}$ (+0.130 \AA) for the $P$(1) blend (Bakker \& 
Lambert 1998). The generally good agreement between the laboratory and 
astronomical results serves as confirmation that our detections of the 
$^{13}$CN $R$(1) and $P$(1) lines are real. 

\section{ISOTOPOLOGIC RATIOS IN C-BEARING MOLECULES}

\subsection{$^{12}$CH$^+$/$^{13}$CH$^+$ Ratios}
One of the primary motivations for this study was to evaluate whether the 
reported variations in the $^{12}$CH$^+$/$^{13}$CH$^+$ ratio toward stars in the 
solar neighborhood are indicative of intrinsic scatter in the local 
interstellar $^{12}$C/$^{13}$C ratio or result instead from systematic effects 
in the individual investigations. The results of our detailed analysis 
suggest that the latter scenario is more likely. In particular, we do not 
confirm the low ratios that were obtained by Hawkins \& Jura (1987) from Lick 
Observatory data toward the Pleiades stars 20~Tau and 23~Tau ($40\pm9$ and 
$41\pm9$, respectively), nor an earlier result toward 20~Tau ($49^{+12}_{-8}$) 
by Vanden Bout \& Snell (1980). Hawkins et al. (1985, 1989) reported low 
$^{12}$CH$^+$/$^{13}$CH$^+$ ratios based on Lick data toward $\zeta$~Oph, as 
well, giving values of $43\pm6$ (Hawkins et al. 1985) and $49\pm10$ (Hawkins et 
al. 1989). However, many subsequent investigations revealed 
$^{12}$CH$^+$/$^{13}$CH$^+$ ratios closer to 70 for this line of sight, 
including Hawkins et al. (1993), who examined Kitt Peak National Observatory 
(KPNO) spectra of $\zeta$~Oph, arriving at a ratio of $63\pm8$. More precise 
results are available from Crane et al. (1991), who used the 2.7 m telescope at 
McDonald Observatory to derive a $^{12}$CH$^+$/$^{13}$CH$^+$ ratio of 
$67.6\pm4.5$, and Stahl \& Wilson (1992), who reanalyzed ESO observations of 
$\zeta$~Oph (Stahl et al. 1989) and found a value of $71\pm3$. The Crane et al. 
(1991) result is, in fact, very close to the mean value of all of the 
$\zeta$~Oph results reported in the literature. Our new determination from 
McDonald data, while not as precise as the Crane et al. (1991) value, is 
nonetheless consistent with it.

\begin{figure*}[!t]
\centering
\includegraphics[angle=90,width=0.95\textwidth]{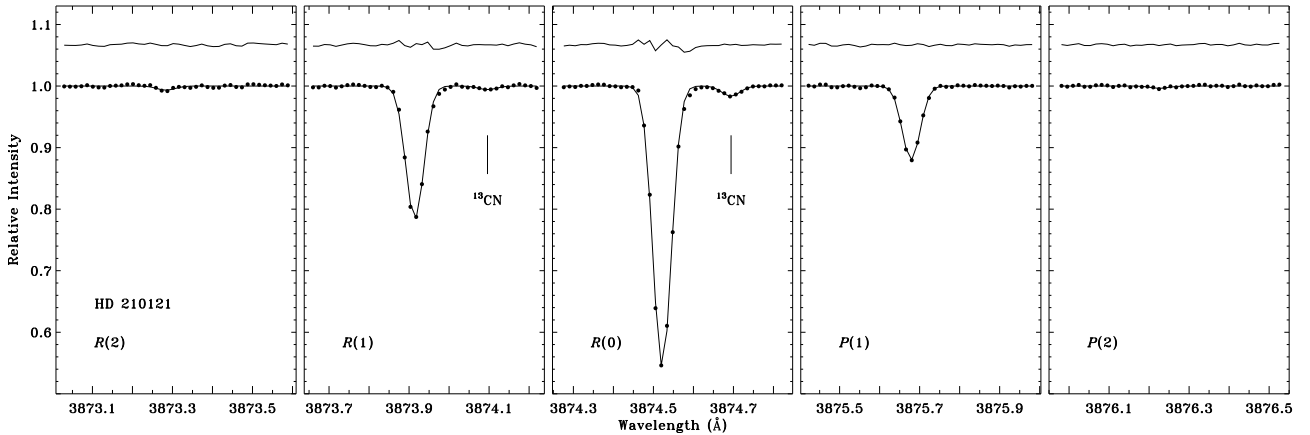}
\caption[Profile synthesis fits to the CN lines toward HD~210121.]{Same as 
Figure 12 except for the CN $B$$-$$X$ (0,~0) band toward HD~210121. The 
positions of the $R$(0) and $R$(1) lines of $^{13}$CN are indicated by tick 
marks.}
\end{figure*}

Hawkins et al. (1993) believe that the discrepancies between their KPNO results 
and the earlier Lick results for $\zeta$~Oph (e.g., Hawkins et al. 1985) were 
caused by telluric contamination of the Lick spectra, which complicated the 
placement of the continuum. While this explanation may have resolved the 
controversy surrounding the $^{12}$CH$^+$/$^{13}$CH$^+$ ratio toward 
$\zeta$~Oph, it left open the possibility that the spectra obtained by Hawkins 
\& Jura (1987), including for 20~Tau and 23~Tau, were not similarly 
contaminated and that the low $^{12}$CH$^+$/$^{13}$CH$^+$ ratios in these 
directions would later be confirmed. With our new results on 
$^{12}$CH$^+$/$^{13}$CH$^+$ toward 20~Tau and 23~Tau, such a position would now 
be difficult to support. The Stahl et al. (2008) value for 
$^{12}$CH$^+$/$^{13}$CH$^+$ in the direction of 23~Tau ($96.5\pm16.3$) is 
further confirmation that a ratio closer to 40 in the Pleiades clouds is 
unlikely, although the UVES results have their own complications (see \S{} 3). 
Initially, we suspected that the difference between our McDonald results and 
those of Hawkins \& Jura (1987) for the Pleiades stars was related to the 
smaller $b$-values we derive when modeling the CH$^+$ profiles. However, our 
optical depth corrections for these sight lines, as determined by the profile 
fitting routine, would only amount to increases of about 4\% in 
$N$($^{12}$CH$^+$) over the column densities obtained by Hawkins \& Jura (1987) 
from their adopted corrections, whereas the $^{12}$CH$^+$/$^{13}$CH$^+$ ratios 
differ by a factor of 2. The dominant cause of the discrepancy is the almost 
factor-of-2 larger equivalent widths that Hawkins \& Jura (1987) find for the 
$^{13}$CH$^+$ lines (see Table 2), which again points to possible errors in 
continuum normalization or to the effects of blending in their more moderate 
resolution spectra.

Federman et al. (2003) analyzed McDonald 2.7 m data toward $\rho$~Oph~A and 
found a $^{12}$CH$^+$/$^{13}$CH$^+$ ratio of $120\pm54$, which, despite the 
large value, is within 1 $\sigma$ of our more precise determination. These 
authors achieved a S/N ratio per pixel of about 650 in the vicinity of CH$^+$ 
$\lambda$4232, whereas the S/N is about 950 per pixel in our $\rho$~Oph~A data. 
As our spectra were also sampled at slightly higher resolution, this case 
illustrates the general trend that higher-resolution, higher signal-to-noise 
observations tend to result in $^{12}$CH$^+$/$^{13}$CH$^+$ ratios that are 
closer to the average interstellar value of $^{12}$C/$^{13}$C. Indeed, our 
McDonald observations of 20~Tau, 23~Tau, $\rho$~Oph~A, and $\zeta$~Oph yield 
$^{12}$CH$^+$/$^{13}$CH$^+$ ratios that are indistinguishable from 70 in each 
case. The weighted mean of the four determinations is $74.4\pm7.6$, or only 
0.6~$\sigma$ higher than the $^{12}$C/$^{13}$C ratio adopted for the local 
interstellar medium. While this mean represents only a small sample of 
interstellar sight lines, the result is significant in the sense that each of 
these sight lines at one time had been found to have a much lower or much 
higher $^{12}$CH$^+$/$^{13}$CH$^+$ ratio. The statistical dispersion (i.e., 
standard deviation) in our new determinations of $^{12}$CH$^+$/$^{13}$CH$^+$ for 
these four sight lines (7.3) is essentially identical to the error in the 
weighted mean (7.6). We interpret these findings as confirmation of the 
theoretical expectation that $^{12}$CH$^+$/$^{13}$CH$^+$ reflects the ambient 
carbon isotopic ratio, and that this ratio does not vary within the solar 
neighborhood.

\subsection{$^{12}$CN/$^{13}$CN Ratios}
The $^{12}$CN/$^{13}$CN ratios presented in Table 7 constitute the most 
extensive isotopologic results to date for CN in diffuse molecular clouds. The 
UVES results, in particular, are considerably more precise than any literature  
results currently available. Gredel et al. (1991) reported abundance ratios of 
$^{12}$CN to $^{13}$CN for four interstellar sight lines. These included 
HD~73882 and HD~169454, for which they found $^{12}$CN/$^{13}$CN ratios of 
$170\pm120$ and $38\pm16$, respectively. Due to the large uncertainties, these 
authors were unable to conclude whether or not the scatter in their sample 
represented true cloud-to-cloud variations. Our new results for these lines of 
sight reveal $^{12}$CN/$^{13}$CN ratios closer to the ambient value of 
$^{12}$C/$^{13}$C in both instances. However, the small uncertainties in our 
determinations allow even modest variations in the relative abundances of 
$^{12}$CN and $^{13}$CN to be discerned. Thus, the moderate increase in 
$^{12}$CN/$^{13}$CN that we find toward HD~73882 likely indicates an actual 
deficit in $^{13}$CN, relative to $^{12}$CN, for the gas in this direction. 
Further confirmation of our result comes from an independent assessment by D. 
Welty (private communication), who finds a $^{12}$CN/$^{13}$CN ratio of 87 for 
the line of sight to HD~73882. For HD~154368, Palazzi et al. (1990) reported a 
$^{12}$CN/$^{13}$CN ratio of $101\pm12$, whereas we find a value essentially 
equal to 70. In this case, and in the comparison with the Gredel et al. (1991) 
results, the discrepancy in the $^{12}$CN/$^{13}$CN ratio is largely driven by 
differences in the measured equivalent width of the $^{13}$CN line (see 
Table 3). However, the Gredel et al. (1991) values of $W_{\lambda}$($^{13}$CN) 
only differ from ours at the 1-$\sigma$ level, because of their significantly 
larger uncertainties, whereas the Palazzi et al. (1990) measurement of 
$W_{\lambda}$($^{13}$CN) for HD~154368 ($0.82\pm0.10$ m\AA) is lower than our 
more precise result ($1.13\pm0.04$ m\AA) by 3 $\sigma$. Equivalent width 
measurements of very weak lines are particularly sensitive to the placement of 
the continuum. However, by examining the raw (unnormalized) spectrum of 
HD~154368, we can find no acceptable continuum fit that would yield a value of 
$W_{\lambda}$($^{13}$CN) as low as the Palazzi et al. (1990) result.

Of the McDonald sight lines, both $\zeta$~Oph and $\zeta$~Per have previously 
been investigated for $^{12}$CN/$^{13}$CN. The most precise result available is 
that of Crane \& Hegyi (1988), who found a $^{12}$CN/$^{13}$CN ratio of 
47.3$^{+5.5}_{-4.4}$ in the direction of $\zeta$~Oph. Later, Roth \& Meyer 
(1995) obtained a ratio of $35\pm13$ for this line of sight. Our new 
determination ($48.8\pm19.5$) has a precision comparable to the Roth \& Meyer 
(1995) value, yet agrees well with the result of Crane \& Hegyi (1988). For 
$\zeta$~Per, Kaiser et al. (1991) found a $^{12}$CN/$^{13}$CN ratio of 
77$^{+27}_{-18}$, which is consistent with the slightly less precise value we 
derive from our McDonald observations of this star ($67.0\pm28.0$). In order to 
obtain a more precise result for $\zeta$~Per, we can combine our determination 
with that of Kaiser et al. (1991) to arrive at a weighted mean 
$^{12}$CN/$^{13}$CN ratio of $72\pm19$. Thus, again, we find a value of 
$^{12}$CN/$^{13}$CN that is indistinguishable from the ambient $^{12}$C/$^{13}$C 
ratio. This is in contrast to the collection of $\zeta$~Oph results, which 
provides strong evidence for a $^{12}$CN/$^{13}$CN ratio that is significantly 
lower than the ambient value. We have already noted the elevated 
$^{12}$CN/$^{13}$CN ratio we find in the direction of HD~73882, which is 
significant because of the small uncertainty. Meyer et al. (1989) find a 
$^{12}$CN/$^{13}$CN ratio of $122\pm33$ toward HD~21483, which, despite the 
larger uncertainty, also suggests a higher-than-ambient value. Together, these 
results indicate that, unlike CH$^+$, CN is affected by fractionation in at 
least some diffuse environments.

The weighted mean of all of our determinations of $^{12}$CN/$^{13}$CN toward 
McDonald and UVES targets is $67.5\pm1.0$, which agrees well with the adopted 
value for the local interstellar $^{12}$C/$^{13}$C ratio. However, there is 
significant scatter in $^{12}$CN/$^{13}$CN from one sight line to another and 
the relatively small uncertainties in our determinations suggest that the 
variation represents true differences among the environments being probed. From 
our sample of $^{12}$CN/$^{13}$CN ratios, we find a dispersion of 24.9, which, 
in contrast to the situation for CH$^+$, is much greater than the error in the 
weighted mean (1.0). These results for CN are similar, statistically, to 
published results for CO. For example, the weighted mean value of 
$^{12}$CO/$^{13}$CO from the ratios tabulated by Sheffer et al. (2007) is 
$69.9\pm1.9$, while the dispersion in those measurements is 29.5. Again, this 
suggests a level of fractionation in CN (and CO) not observed in CH$^+$. 
However, some caution on this point may be warranted. First, we are unable to 
determine reliable $^{12}$CH$^+$/$^{13}$CH$^+$ ratios for the majority of the 
UVES sight lines, meaning that our sample of CH$^+$ ratios is considerably 
smaller than that for CN. Second, the CN ratios toward HD~161056 and HD~170740, 
the two sight lines contributing the most to the scatter in $^{12}$CN/$^{13}$CN, 
are somewhat uncertain. The $^{12}$CN and $^{13}$CN features are blended in the 
spectrum of HD~161056 because of the velocity separation between the two 
line-of-sight components. This may also contribute to the somewhat higher 
isotope shift we find in this direction. For HD~170740, there is some 
uncertainty in the placement of the continuum due to what may be either 
telluric absorption or the result of instrumental fringing just to the red of 
the $^{13}$CN feature, which is also quite weak. Without these two extreme 
results, the scatter in $^{12}$CN/$^{13}$CN is reduced to 10.3, though this 
might still be significant given the small error in the weighted mean value.

\begin{figure}[!t]
\centering
\includegraphics[width=0.45\textwidth]{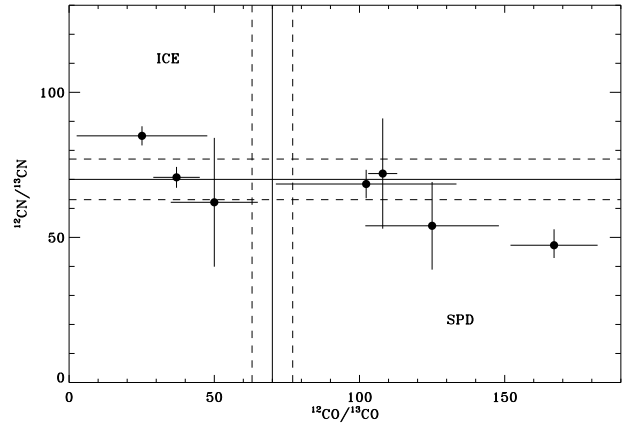}
\caption[$^{12}$CN/$^{13}$CN vs. $^{12}$CO/$^{13}$CO.]{$^{12}$CN/$^{13}$CN versus 
$^{12}$CO/$^{13}$CO for the seven sight lines where both ratios have been 
determined. The solid and dashed lines denote the mean and 1-$\sigma$ 
deviations of the local interstellar $^{12}$C/$^{13}$C ratio ($70\pm7$; see 
Sheffer et al. 2007). Data points in the upper left quadrant are influenced by 
isotopic charge exchange (ICE), while those in the lower right are dominated by 
selective photodissociation (SPD).}
\end{figure}

\subsection{$^{12}$CN/$^{13}$CN versus $^{12}$CO/$^{13}$CO}
Many of the targets selected for this investigation were chosen because they 
exhibit fractionated values of $^{12}$CO/$^{13}$CO (see Table 7). That is to say 
their $^{12}$CO/$^{13}$CO ratios are significantly above or below the ambient 
value of $^{12}$C/$^{13}$C. Since CN and CO are presumed to reside in the same 
portion of a diffuse molecular cloud (Pan et al. 2005), the expectation is that 
CN will be fractionated in the opposite sense compared to CO. This is due to 
the fact that CO is the most abundant C-bearing molecule in the ISM (by 2$-$3 
orders of magnitude over CN), and should regulate the availability of $^{13}$C 
atoms for incorporation into other C-bearing molecules that are cospatial and 
must draw from the same carbon reservoir. Figure 19 plots $^{12}$CN/$^{13}$CN 
versus $^{12}$CO/$^{13}$CO for the seven sight lines where both ratios have been 
determined. The $^{12}$CN/$^{13}$CN ratios are from this investigation (Table 7) 
except in the case of $\zeta$~Oph, where we have used the more precise 
determination by Crane \& Hegyi (1988), and in the case of $\zeta$~Per, where 
we have plotted the weighted mean of our result and that of Kaiser et al. 
(1991). The $^{12}$CO/$^{13}$CO ratios are either from our archival study using 
\emph{HST} (Sheffer et al. 2007) or from the literature (Hanson et al. 1992; 
Sonnentrucker et al. 2007). Also indicated in the figure is the value we adopt 
for the local interstellar $^{12}$C/$^{13}$C ratio ($70\pm7$; see Sheffer et al. 
2007), which divides the parameter space into four quadrants. Two of these 
quadrants are of particular interest: the upper left quadrant, where low values 
of $^{12}$CO/$^{13}$CO are associated with high values of $^{12}$CN/$^{13}$CN, 
and the lower right quadrant, where high $^{12}$CO/$^{13}$CO and low 
$^{12}$CN/$^{13}$CN ratios coexist. In molecular gas characterized by a low 
$^{12}$CO/$^{13}$CO ratio, isotopic charge exchange has presumably converted 
some of the $^{12}$CO into $^{13}$CO, while also exchanging $^{13}$C$^+$ ions 
for $^{12}$C$^+$. Since the formation of CN is initiated by reactions involving 
C$^+$ (e.g., Federman et al. 1994), the gas in an environment where CO is 
fractionated by ICE should be enhanced in $^{12}$CN relative to $^{13}$CN. In 
contrast, a high $^{12}$CO/$^{13}$CO ratio indicates that the molecular gas has 
been subjected to selective photodissociation, which preferentially destroys 
$^{13}$CO when only $^{12}$CO is self shielded. In regions where SPD is 
occurring, the dissociated $^{13}$C atoms are quickly ionized and incorporated 
into the pathways that lead to CN, which should result in enhanced $^{13}$CN 
production.

The data plotted in Figure 19 are suggestive of the inverse relationship 
between $^{12}$CN/$^{13}$CN and $^{12}$CO/$^{13}$CO that is expected due to 
photochemical fractionation and the physical association of CN and CO in 
diffuse cloud cores. Virtually no data points populate the upper right or lower 
left quadrants of the figure where the fractionation of CN and CO occurs in the 
same directions. The two data points that do (just barely) lie in these regions 
are for the sight lines to 20~Aql (in the lower left) and $\zeta$~Per (in the 
upper right). These sight lines have large uncertainties associated with their 
$^{12}$CN/$^{13}$CN ratios, which in both cases are consistent with the 
unfractionated value. Although the downward trend in $^{12}$CN/$^{13}$CN with 
increasing $^{12}$CO/$^{13}$CO is apparent, the correlation between these two 
parameters is rather shallow. In addition, the appearance of a trend is 
somewhat contingent upon the data points at the extremes. The sight line with 
the lowest $^{12}$CO/$^{13}$CO ratio (and highest value of $^{12}$CN/$^{13}$CN) 
is HD~73882, while the highest values of $^{12}$CO/$^{13}$CO (and lowest 
$^{12}$CN/$^{13}$CN ratios) are found in the directions of $\rho$~Oph~A and 
$\zeta$~Oph (in order of increasing $^{12}$CO/$^{13}$CO). All of the 
$^{12}$CN/$^{13}$CN ratios between these extremes are consistent with the 
ambient value. Yet the uncertainties in $^{12}$CN/$^{13}$CN toward HD~73882 and 
$\zeta$~Oph are among the smallest in the sample, giving credence to the 
perceived trend.

The severity of CN fractionation depends not only on the effects that alter the 
relative abundances of $^{12}$CO and $^{13}$CO but also on the balance between 
CO and C$^+$, the other major reservoir of carbon in diffuse gas. If CO and 
C$^+$ have roughly equivalent abundances, then any alteration in 
$^{12}$CO/$^{13}$CO will have an equal, but opposite, effect on 
$^{12}$C$^+$/$^{13}$C$^+$ and hence on $^{12}$CN/$^{13}$CN. If C$^+$ dominates 
over CO, however, fractionation in CO will have less of an impact on 
$^{12}$CN/$^{13}$CN because there will be a larger pool of C$^+$ ions that will 
maintain the ambient 12-to-13 ratio. The reverse is also true. When CO 
dominates, CO fractionation will have a greater influence on the ratio of 
$^{12}$CN to $^{13}$CN. A detailed modeling effort to track how the relative 
abundance of each isotope (and isotopologue) of interest depends on the 
physical conditions in the cloud and the parameters of the external UV 
radiation field is beyond the scope of this paper. Instead, we employ a simple, 
analytical method to estimate the fraction of carbon that is locked up in CO 
along a given line of sight based on the measured isotopologic ratios in CN and 
CO. In what follows, we define $f(^{13}$CN$) = 
N(^{13}$CN$)/[N(^{12}$CN$)+N(^{13}$CN$)] = 1/[N(^{12}$CN$)/N(^{13}$CN$)+1]$ as 
the fraction of CN molecules that take the form of $^{13}$CN. Similar 
definitions can be made for $^{13}$CO and $^{13}$CH$^+$, where 
$f(^{13}$CH$^+) \equiv f(^{13}$C$) = 1/(70+1) = 0.014$ is taken to be the 
ambient fraction of $^{13}$C. With these definitions, we can write 
$f$($^{13}$CN) as a function of $f$($^{13}$CO) and $f$($^{13}$CH$^+$). We have

\begin{eqnarray}
f(^{13}\mathrm{CN}) & = & f(^{13}\mathrm{CH}^+)[1-x] \nonumber \\
 &   & +\left\{f(^{13}\mathrm{CH}^+)+\left[f(^{13}\mathrm{CH}^+)
-f(^{13}\mathrm{CO})\right]\right\}x \nonumber \\
 & = & f(^{13}\mathrm{CH}^+)+\left[f(^{13}\mathrm{CH}^+)
-f(^{13}\mathrm{CO})\right]x,
\end{eqnarray}

\noindent
where the factor $x$ determines how strongly $f$($^{13}$CN) is affected by 
changes in $f$($^{13}$CO) and is equivalent to the column density ratio 
$N$(CO)/$N$(C$^+$), assuming the abundance of neutral carbon is negligible. In 
the limiting case $x=0$ [i.e., $N$(C$^+$) $\gg$ $N$(CO)], CN is unaffected by 
fractionation in CO and $^{12}$CN/$^{13}$CN equals the ambient ratio.

Equation (1) can be rearranged to yield an expression for $x$,

\begin{equation}
x=\frac{f(^{13}\mathrm{CN})-f(^{13}\mathrm{CH}^+)}{f(^{13}\mathrm{CH}^+)
-f(^{13}\mathrm{CO})},
\end{equation}

\noindent
which is valid whether CO is fractionated upward via selective 
photodissociation or downward due to isotopic charge exchange. However, when 
selective photodissociation is under consideration, the fractions should be 
defined as, for example, $f(^{13}$CN$) = N(^{13}$CN$)/N(^{12}$CN$)$, because SPD 
does not exchange one isotopologue for another but simply destroys the less 
abundant isotopologue leaving the more abundant one unchanged. The fraction of 
total carbon that is locked up in CO can be found from $x$ by $f($CO$) = 
N($CO$)/[N($CO$)+N($C$^+)]=[N($CO$)/N($C$^+)]/[N($CO$)/N($C$^+)+1]=x/(x+1)$. We 
use this relation in conjunction with equation (2) to derive $f$(CO) for the 
sight lines included in Figure 19. The resulting values are $f($CO$) = 0.09$ 
for HD~73882, 0.01 for HD~154368, 0.07 for HD~210121, 0.40 for $\rho$~Oph~A, 
and 0.45 for $\zeta$~Oph. No physically meaningful determinations of $f$(CO) 
are possible toward 20~Aql or $\zeta$~Per because the measured isotopologic 
ratios in CN and CO fall on the same side of the ambient ratio. Again, we note 
that in these cases the $^{12}$CN/$^{13}$CN ratios are consistent with being 
unfractionated.

For the sight lines in Ophiuchus, we find that nearly 50\% of the carbon is 
contained in CO, which is consistent with the fact that the UV optical depths 
($\tau_{\mathrm{UV}}$ = 1.98 and 2.04 for the clouds toward $\zeta$~Oph and 
$\rho$~Oph~A, respectively) are near that where the C$^+$ to CO conversion is 
expected to occur (i.e., $\tau_{\mathrm{UV}}=2$; see Federman et al. 1994). The 
value of $f$(CO) is substantially lower toward HD~210121 even though the UV 
optical depth in this direction ($\tau_{\mathrm{UV}}=3.35$; Sheffer et al. 2008) 
is larger than those for the Ophiuchus clouds. Here, we note that the 
$^{12}$CO/$^{13}$CO ratio in the direction of HD~210121 is consistent with 70 
and so does not show strong evidence of fractionation, meaning that $f$(CO) 
cannot be determined. An unfractionated value of $^{12}$CO/$^{13}$CO toward 
HD~210121 is also consistent with the suggestion by Welty \& Fowler (1992) that 
the UV radiation field incident on the cloud in this direction (the 
high-latitude molecular cloud DBB 80; see de Vries \& van Dishoeck 1988) is 
lower than the average interstellar field. Finally, the low values of $f$(CO) 
we find toward HD~73882 and HD~154368 are surprising considering the much 
larger UV optical depths in these directions (adopting 
$\tau_{\mathrm{UV}}\simeq2A_V\simeq4.7$ in both cases). The larger values of 
$\tau_{\mathrm{UV}}$ indicate that the sight lines to HD~73882 and HD~154368 
probe deeper into their respective clouds, where a majority of the carbon is 
expected to be in the form of CO rather than in neutral atomic or ionic form. 
However, equation (2) only applies if the $^{12}$CN/$^{13}$CN ratio reflects the 
$^{12}$C$^+$/$^{13}$C$^+$ ratio of the gas from which CN formed. If 
$^{12}$CN/$^{13}$CN is substantially altered after CN formation, then equation 
(2) no longer holds. Thus, instead of surmising that there is a very low 
fraction of elemental carbon in CO in the lines of sight to HD~73882 and 
HD~154368, we suggest that we are seeing the effects of a separate 
fractionation process in CN, which acts to lower $^{12}$CN/$^{13}$CN such that 
the enhancements due to CO fractionation are not as large as expected.

We assume that the low $^{12}$CO/$^{13}$CO ratios in the directions of HD~73882 
and HD~154368 result from an efficient isotopic charge exchange reaction 
involving CO, which implies a colder kinetic temperature for the molecular gas 
along these lines of sight. Yet, at colder temperatures, CN is similarly 
affected by ICE via the reaction 
$^{13}$C$^+$~+~$^{12}$CN~$\to$~$^{12}$C$^+$~+~$^{13}$CN~+~$\Delta E$, where the 
zero-point energy difference $\Delta E/k_{\mathrm{B}}=31$~K (Bakker \& Lambert 
1998; derived from the molecular parameters in Prasad et al. 1992) favors 
$^{13}$CN. Unlike CO, however, CN is dissociated by continuum radiation, not by 
lines, and so isotope-selective self shielding should not be a factor. Any 
dissociating photons will equally destroy $^{12}$CN as well as $^{13}$CN. Yet if 
the timescale for the ICE reaction with CN is shorter than the 
photodissociation timescale, then the reaction will reach equilibrium and we 
can write

\begin{equation}
\frac{n(^{12}\mathrm{CN})}{n(^{13}\mathrm{CN})}=
\frac{n(^{12}\mathrm{C}^+)}{n(^{13}\mathrm{C}^+)}\frac{k_r}{k_f}=
\frac{n(^{12}\mathrm{C}^+)}{n(^{13}\mathrm{C}^+)}\mathrm{exp}\left(
-\frac{\Delta E}{k_{\mathrm{B}}T_{\mathrm{kin}}}\right),
\end{equation}

\noindent
where $n$(X) represents the number density of species X, $k_f$ is the rate 
coefficient of the forward reaction, and 
$k_r=k_f\mathrm{exp}(-\Delta E/k_{\mathrm{B}}T_{\mathrm{kin}})$ is the rate 
coefficient of the reverse reaction. Analogous expressions have been derived in 
the case of CO (see Sheffer et al. 1992; Lambert et al. 1994). When 
photodissociation of CN dominates over isotopic charge exchange, no upward 
fractionation occurs, as it does for CO, because the photodissociation rates of 
the two CN isotopologues are equal (i.e., $\Gamma_{12}=\Gamma_{13}$) and we have

\begin{equation}
\frac{n(^{12}\mathrm{CN})}{n(^{13}\mathrm{CN})}=
\frac{n(^{12}\mathrm{C}^+)}{n(^{13}\mathrm{C}^+)}\frac{\Gamma_{13}}{\Gamma_{12}}=
\frac{n(^{12}\mathrm{C}^+)}{n(^{13}\mathrm{C}^+)}.
\end{equation}

It is important to note that the $^{12}$C$^+$/$^{13}$C$^+$ ratio in equations 
(3) and (4) refers to the ratio that has been elevated above the ambient value 
of $^{12}$C/$^{13}$C by isotopic charge exchange with CO. Thus, when isotopic 
charge exchange with CN lowers $^{12}$CN/$^{13}$CN, this ratio is still higher 
than the ambient value. If we adopt a kinetic temperature of 51 K for the 
clouds toward HD~73882 and HD~154368, a value which corresponds to the observed 
H$_2$ rotational temperature ($T_{01}$) for both sight lines (Rachford et al. 
2002), then we find, from equation (3), that ICE would have lowered 
$^{12}$CN/$^{13}$CN by a factor of 1.84. If the actual $^{12}$C$^+$/$^{13}$C$^+$ 
ratios toward HD~73882 and HD~154368 are higher than the measured 
$^{12}$CN/$^{13}$CN ratios by this amount, then the values of $f$(CO) in these 
directions would be 0.24 and 0.34, respectively. However, the rotational 
excitation temperature in C$_2$ ($T_{02}$) may provide a better indication of 
the kinetic temperature in the denser central portion of a diffuse cloud where 
CN resides. Indeed, Sonnentrucker et al. (2007) find that $T_{02}$(C$_2$) is 
consistently lower than $T_{01}$(H$_2$) for a given sight line. These authors 
fit theoretical models for the excitation to the observed C$_2$ rotational 
level populations to estimate the actual kinetic temperature, finding 
$T_{\mathrm{kin}}=20\pm5$ K for both HD~73882 and HD~154368. At this kinetic 
temperature, the true $^{12}$C$^+$/$^{13}$C$^+$ ratios would be a factor of 4.71 
higher than $^{12}$CN/$^{13}$CN, resulting in $f$(CO) values of 0.32 for 
HD~73882 and 0.48 for HD~154368. Thus, the inclusion of an isotopic charge 
exchange reaction with CN leads to more realistic values for the fraction of C 
in CO for sight lines with low $^{12}$CO/$^{13}$CO ratios. However, even at 
20~K, the derived value of $f$(CO) toward HD~73882 is smaller than would be 
expected. With our simple analytical model, it is difficult to accommodate a 
$^{12}$CO/$^{13}$CO ratio as low as 25 and also demand that a high fraction of 
carbon be locked up in CO. The uncertainty in the determination of 
$^{12}$CO/$^{13}$CO toward HD~73882 is considerable, however, and indicates that 
a ratio as high as 47 would be consistent with the observations. At 
$T_{\mathrm{kin}}=20$ K, a $^{12}$CO/$^{13}$CO ratio of 47 would result in a value 
of $f($CO$)=0.63$ for HD~73882. In the case of HD~154368, the maximum 
$^{12}$CO/$^{13}$CO ratio allowed by the much smaller reported uncertainty is 
45, which, at $T_{\mathrm{kin}}=20$ K, would yield $f($CO$)=0.59$. As these 
calculations demonstrate, there is a real need for expanding on existing 
chemical models of diffuse clouds in order to clarify the situation regarding 
photochemical fractionation in all of the relevant C-bearing molecules and 
atomic carbon species.

\section{CN ROTATIONAL EXCITATION}
The CN column densities resulting from our profile synthesis fits to the 
$B$$-$$X$ (0,~0) band can be used to derive the CN rotational excitation 
temperatures $T_{01}$ and $T_{12}$ by applying the Boltzmann equation

\begin{equation}
T_{ij}=\frac{h\nu}{k_{\mathrm{B}}}\left[\mathrm{ln}\left(\frac{g_jN_i}{g_iN_j}
\right)\right]^{-1},
\end{equation}

\noindent
where $g$ and $N$ denote the statistical weight and column density of the lower 
($i$th) or upper ($j$th) rotational level. The energy separation associated 
with the $N=1$$-$$0$ transition at $\lambda=2.64$ mm is taken to be 
$h\nu$/$k_{\mathrm{B}}=5.445$~K, while for $N=2$$-$$1$ at $\lambda=1.32$ mm, we 
assume $h\nu$/$k_{\mathrm{B}}=10.885$ K (Black \& van Dishoeck 1991). 
Uncertainties in $T_{ij}$ were calculated from the column density uncertainties 
($\sigma_N$) through standard error propagation such that

\begin{equation}
\sigma_{T_{ij}}=\frac{h\nu}{k_{\mathrm{B}}}\left[\mathrm{ln}\left(
\frac{g_jN_i}{g_iN_j}\right)\right]^{-2}\left[\left(\frac{\sigma_{N_i}}{N_i}
\right)^2+\left(\frac{\sigma_{N_j}}{N_j}\right)^2\right]^{1/2}.
\end{equation}

\noindent
The excitation temperatures and 1-$\sigma$ errors obtained from equations (5) 
and (6) are listed in Table 9, along with the column densities of $^{12}$CN in 
the first three rotational levels of the ground vibrational state. Also shown 
are the values of $T_{01}$ for the $^{13}$CN molecule for the four UVES sight 
lines with detections of $^{13}$CN $R$(1). For HD~73882, HD~154368, and 
HD~210121, the $N=1$ column density of $^{13}$CN is based solely on the value 
obtained from the $R$(1) line. Since both the $R$(1) and $P$(1) lines of 
$^{13}$CN are detected in the spectrum of HD~169454, the final value of 
$N(N=1)$ for $^{13}$CN in this direction is the weighted mean of the column 
densities derived from the two transitions.

As expected, our derived rotational temperatures are consistent with the 
dominant excitation mechanism being radiative pumping from the cosmic microwave 
background (see Figure 20). The weighted mean value of $T_{01}$($^{12}$CN) in 
our sample is $2.754\pm0.002$ K, which implies an excess over the temperature 
of the CMB ($T_{\mathrm{CMB}}=2.725\pm0.002$ K; Mather et al. 1999) of only 
$29\pm3$~mK. This modest excess likely arises from local collisional excitation 
by electrons in some environments (see below). However, the dispersion in 
individual values of $T_{01}$ (134 mK) cautions that the excess may not be 
physical but rather due to measurement or modeling uncertainties in our 
analysis. There is a slight indication of a greater excess in 
$T_{12}$($^{12}$CN) based on the weighted mean of $2.847\pm0.014$ K, but the 
dispersion in these measurements (262 mK) is also greater. An elevated value of 
$T_{12}$ could be caused by an overestimation of the strength of the weak 
$R$(2) and $P$(2) lines, resulting from noise in the spectrum. This is likely 
to blame for the higher values of $T_{12}$ we find toward the McDonald targets, 
as well as toward HD~161056 and HD~170740. These are the sight lines with the 
weakest CN absorption in our sample. For the directions where CN absorption is 
the strongest, namely, HD~73882, HD~154368, HD~169454, and HD~210121, the 
agreement between $T_{01}$ and $T_{12}$ is at the 3\% level or better. The 
rotational excitation temperatures derived from the observed $^{13}$CN $R$(0), 
$R$(1), and $P$(1) lines show no excess over the temperature of the CMB. 
Quantitatively, the weighted mean value of $T_{01}$($^{13}$CN) is 
$2.652\pm0.069$ K, which is within 1 $\sigma$ of $T_{\mathrm{CMB}}$.

Recently, a compilation of CN excitation temperatures along over 100 lines of 
sight was published by S\l{}yk et al. (2008). These authors claim to detect an 
excess over the cosmic microwave background of 580~mK based on their average 
temperature of 3.31 K. Upon closer inspection, however, the rather high values 
of $T_{01}$ given in S\l{}yk et al. (2008) are almost certainly the result of 
not accounting adequately for optical depth effects in the strong $R$(0) line. 
While the authors make numerous attempts at estimating the Doppler widths, they 
settle on a simple assumption that $b=1.0$ km s$^{-1}$ for all directions. In 
our much smaller sample, the vast majority of CN components have $b<1.0$ 
km~s$^{-1}$ and we note that differences of only a few tenths of a km s$^{-1}$ 
can greatly impact the derived column densities, even for lines with moderate 
optical depths. If we restrict the S\l{}yk et al. (2008) sample to only the 
eleven sight lines in common with our investigation, then their average 
excitation temperature is still 3.30 K. The largest discrepancies between the 
two samples arise for the three sight lines with the most severe cases of 
optically thick absorption, HD~73882, HD~154368, and HD~169454, for which 
S\l{}yk et al. (2008) give $T_{01}=3.68$ K, $3.67$ K, and $5.06$~K, 
respectively. In each of these cases, their measured values of $W_{\lambda}$ 
agree with ours at the 5\% level. However, their column densities, especially 
those derived from the $R$(0) line, are significantly smaller than ours [by 
40\% for the $R$(0) lines toward HD~73882 and HD~154368 and 76\% in the case of 
$R$(0) toward HD~169454]. The smaller values S\l{}yk et al. infer for $N(N=0)$ 
are then directly responsible for their higher values of $T_{01}$.

Palazzi et al. (1992) previously examined CN rotational excitation in a sample 
of 34 interstellar sight lines, ten of which are included in our investigation. 
Their weighted mean of $T_{01}=2.817\pm0.022$ K is within 3~$\sigma$ of our 
more precise result. From their mean value, Palazzi et al. (1992) deduced an 
excess temperature of $82\pm30$~mK, which would be 92 mK relative to the value 
of $T_{\mathrm{CMB}}$ adopted here. Again, if we consider only the ten sight 
lines from Palazzi et al. (1992) that are also analyzed in the present work, 
their average excitation temperature is 2.813 K, which is essentially identical 
to their full sample mean. Evidently, sample size has little impact on the mean 
value of $T_{01}$($^{12}$CN), which is more heavily dependent on systematic 
effects, and, specifically, on corrections for optical depth. The basic 
difference between our study and that of Palazzi et al. (1992) is that, for the 
ten sight lines in common, Palazzi et al. find $b$-values that are, on average, 
0.1 km s$^{-1}$ higher than the ones we derive. This is a minor difference, but 
it accounts for the 2\% higher mean excitation temperature that these authors 
obtained. We also note that, for this sample of ten sight lines, we have 
reduced the scatter in $T_{01}$ from 0.3 K to 0.1 K.

\begin{figure}[!t]
\centering
\includegraphics[width=0.45\textwidth]{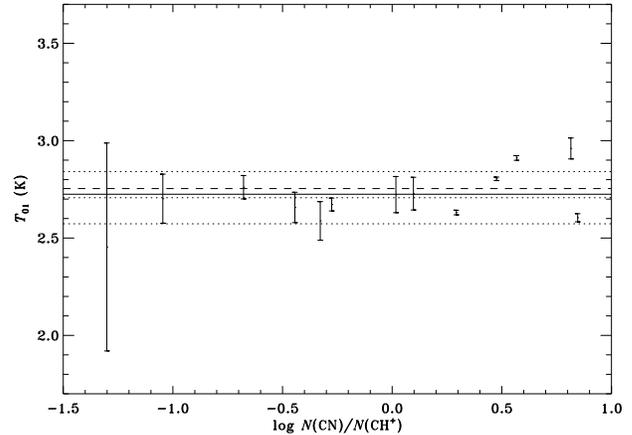}
\caption[CN excitation temperatures versus the CN/CH$^+$ column density ratio.]
{Rotational excitation temperatures in $^{12}$CN versus the CN/CH$^+$ column 
density ratio for individual velocity components. The full 3-$\sigma$ range in 
$T_{01}$($^{12}$CN) is shown in each case. The solid line denotes the 
temperature of the Cosmic Microwave Background ($T_{\mathrm{CMB}}=2.725$~K; 
Mather et al. 1999), while the dashed line represents the weighted mean value 
of $T_{01}$ for our sample ($2.754$~K). The dotted lines indicate the 
(unweighted) mean and standard deviation for this set of measurements 
($2.707\pm0.134$~K). The two data points that lie above the 1-$\sigma$ range 
for the sample are for the main component toward HD~154368 and for the 
$-$3.1~km~s$^{-1}$ component toward HD~161056 (in order of increasing 
CN/CH$^+$).}
\end{figure}

Although our mean CN excitation temperature ($2.754\pm0.002$ K) implies a 
smaller excess over the temperature of the CMB ($29\pm3$ mK) than has been 
found previously, our result has greater statistical significance because of 
the reduced uncertainties. However, two of the four sight lines in our sample 
with the strongest CN absorption lines, and therefore the smallest errors in 
$T_{01}$, have excitation temperatures below $T_{\mathrm{CMB}}$. These are the 
sight lines to HD~73882, where $T_{01}=2.631\pm0.004$ K (3.4\% below 
$T_{\mathrm{CMB}}$), and HD~210121, for which $T_{01}=2.604\pm0.007$ K (4.4\% 
below $T_{\mathrm{CMB}}$). The rotational excitation temperatures in the other 
two directions with strong CN are significantly above $T_{\mathrm{CMB}}$. For 
HD~154368, we find $T_{01}=2.911\pm0.004$ K (6.8\% above $T_{\mathrm{CMB}}$), 
while for the sight line to HD~169454, we obtain $T_{01}=2.804\pm0.003$ K 
(2.9\% above $T_{\mathrm{CMB}}$). Since there is no known physical mechanism that 
can lower $T_{01}$ below the temperature of the cosmic microwave background, 
our results for HD~73882 and HD~210121 indicate that perhaps the uncertainties 
have been underestimated in these instances. The sight lines to HD~73882, 
HD~154368, HD~169454, and HD~210121 exhibit the largest CN optical depths in 
our sample, meaning that the excitation temperatures are sensitive to small 
changes in $b$. Increases in $b$ of only 0.03 km s$^{-1}$ and 0.06 km~s$^{-1}$ 
for HD~73882 and HD~210121, respectively, would be sufficient to bring the 
derived values of $T_{01}$ into agreement with $T_{\mathrm{CMB}}$. (Likewise, for 
HD~154368 and HD~169454, decreases in $b$ of 0.04 km s$^{-1}$ and 0.01 
km~s$^{-1}$, respectively, would eliminate the small excesses in $T_{01}$ 
observed in these directions.) Such small changes would not be unreasonable 
considering the level of variation present in the $b$-values obtained through 
the doublet ratio method for these sight lines (see Table~5).

Figure~20 plots our derived CN excitation temperatures for individual velocity 
components (shown as 3-$\sigma$ error bars) against the CN/CH$^+$ column 
density ratio, an empirical density indicator (see Sheffer et al. 2008). The 
CMB temperature ($2.725\pm0.002$ K; Mather et al. 1999) and our weighted mean 
value ($2.754\pm0.002$ K) are indicated in the figure. Also shown are the 
(unweighted) mean and standard deviation for our sample of excitation 
temperatures ($2.707\pm0.134$ K). In all but two cases, the derived values of 
$T_{01}$ are within one standard deviation of the unweighted mean, even if the 
individual errors are much smaller. (The small error bars at high density that 
are below $T_{\mathrm{CMB}}$ are for the sight lines to HD~73882 and HD~210121 
discussed above.) Only in the main component toward HD~154368 and in the 
$-$3.1~km~s$^{-1}$ component toward HD~161056 is $T_{01}$ more than 134 mK 
larger than the average and both of these clouds have high CN/CH$^+$ ratios. 
For the latter component, we find $T_{01}=2.960\pm0.018$~K. This result may be 
particularly significant since the CN column densities toward HD~161056 seem 
not to be affected by optical depth effects in any appreciable way. 
Additionally, the excitation temperature in the +2.8 km s$^{-1}$ component in 
this direction ($T_{01}=2.761\pm0.020$ K) is consistent with $T_{\mathrm{CMB}}$ 
at the 2-$\sigma$ level and this component has a CN/CH$^+$ ratio that is a 
factor 30 lower than the ratio in the $-$3.1~km~s$^{-1}$ component. A 
rotational temperature in excess of $T_{\mathrm{CMB}}$ results from a local 
excitation mechanism, presumably electron impact (Thaddeus 1972). As such, an 
excess in $T_{01}$ might be expected at higher gas densities, if the electron 
density in the gas is also higher.

To further explore the local excitation process, we can use our determinations 
of $T_{01}$ to derive estimates for the electron density $n_e$ in these clouds. 
Our approach differs from previous analyses of this kind (e.g., Meyer \& Jura 
1985; Palazzi et al. 1992; Roth \& Meyer 1995) in that past investigators have 
used an independent measure of the contribution to $T_{01}$ from local 
collisional excitation, $T_{\mathrm{loc}}$, to correct their $T_{01}$ values in 
order to derive the temperature of the CMB. These studies have estimated the 
local contribution either from ionization balance by using column densities of 
Ca~{\small I} and Ca~{\small II}, for example, to calculate $n_e$, or from 
microwave observations of CN emission at 2.64 mm to evaluate $T_{\mathrm{loc}}$ 
directly. In our analysis, we adopt $T_{\mathrm{CMB}}=2.725$ K, and assume that 
any additional excitation results from electron collisions. The excess 
temperature, then, yields an estimate for the electron density in the gas. 
Following Thaddeus (1972), the rate of excitation from $N=0$ to $N=1$ may be 
written

\begin{eqnarray}
C_{01} & = & \left(\frac{g_1}{g_0}\right)A_1\mathrm{exp}\left(-
\frac{h\nu}{k_{\mathrm{B}}T_{01}}\right)\left[1-\mathrm{exp}\left(-
\frac{h\nu}{k_{\mathrm{B}}T_{01}}\right)\right]^{-1} \nonumber \\
 & = & \left(\frac{g_1}{g_0}\right)A_1\left[\mathrm{exp}\left(
\frac{h\nu}{k_{\mathrm{B}}T_{01}}\right)-1\right]^{-1},
\end{eqnarray}

\noindent
where $A_1=1.19\times10^{-5}$ s$^{-1}$ (Thaddeus 1972) is the spontaneous 
radiative decay rate of the $N=1$ level and $C_{01}$ represents the total rate, 
including both radiative and collisional excitation. If 
$T_{01}=T_{\mathrm{CMB}}$, then the excitation rate is 
$C_{01}=C_{\mathrm{CMB}}=5.60\times10^{-6}$ s$^{-1}$. Any local excitation will 
increase this rate. Thus, when $T_{01}>T_{\mathrm{CMB}}$, the local rate is given 
by $C_{\mathrm{loc}}=C_{01}-C_{\mathrm{CMB}}$. Once $C_{\mathrm{loc}}$ is known, the 
electron density can be found from $n_e=C_{\mathrm{loc}}/r_{01}$, where $r_{01}$ 
is the rate coefficient for CN rotational excitation by electron impact. 

We calculated $n_e$ from equation (7) and the above relations for clouds with 
significant excesses in $T_{01}$ by assuming a gas kinetic temperature of 20 K. 
This temperature corresponds to that calculated by Sonnentrucker et al. (2007) 
for the clouds toward HD~154368 and HD~169454 from their analysis of C$_2$ 
rotational excitation. The rate coefficient given by Allison \& Dalgarno (1971) 
for CN excitation by electron impact at $T_{\mathrm{kin}}=20$~K is 
$r_{01}=1.3\times10^{-6}$ cm$^3$ s$^{-1}$. For the main component toward 
HD~154368, which has an excess rotational temperature of 186 mK, these 
calculations yield an electron density of 0.69 cm$^{-3}$. For the $-$3.1 
km~s$^{-1}$ component toward HD~161056, where the measured excess is 235 mK, we 
find $n_e$ = 0.88 cm$^{-3}$. Finally, for HD~169454, the excess of 79 mK gives 
$n_e$ = 0.29 cm$^{-3}$. These electron densities are generally larger than 
those derived by Black \& van Dishoeck (1991), who found values of $n_e$ in the 
range 0.02$-$0.5 cm$^{-3}$, and would indicate large gas densities in the cloud 
cores ($n_{\mathrm{H}}\sim6000$ cm$^{-3}$ for the component at $-$3.1 km~s$^{-1}$ 
toward HD~161056, assuming $n_e/n_{\mathrm{H}}\sim1.4\times10^{-4}$). The problem 
is even more severe in cases where $f$(CO) is substantial [e.g., toward 
HD~154368, where we find $f($CO$)\sim0.5$], since fewer carbon atoms are 
available to provide free electrons. For comparison, Sonnentrucker et al. 
(2007) find $n_{\mathrm{H}}\sim250$~cm$^{-3}$ for HD~154368 and HD~169454 based 
on their C$_2$ analysis, while Sheffer et al. (2008) obtain 
$n_{\mathrm{H}}\sim750$~cm$^{-3}$ for the main component toward HD~154368 using 
an analytical expression involving the formation and destruction pathways for 
CN.

It is probably more appropriate to consider the electron densities derived from 
the excess rotational temperatures in CN to be upper limits because excitation 
from neutral impacts (mainly H$_2$) can also contribute. Moreover, the actual 
excess temperature due to local collisional excitation is likely to be less 
than the measured value of $T_{\mathrm{loc}}=T_{01}-T_{\mathrm{CMB}}$, considering 
the dispersion of 134 mK in our determinations of $T_{01}$. Taking this 
dispersion into account, the actual excess temperatures for the clouds toward 
HD~154368 and HD~161056 might be as low as 52 mK and 101 mK, respectively. 
Indeed, from observations of CN emission at 2.64 mm, Palazzi et al. (1990) find 
$T_{\mathrm{loc}}=35\pm9$ mK for the main component toward HD~154368. These lower 
excesses would yield electron densities of $n_e$ = 0.19 cm$^{-3}$ for HD~154368 
and 0.37 cm$^{-3}$ for the $-$3.1 km s$^{-1}$ component toward HD~161056. An 
excess in $T_{01}$ would then not be inferred for HD~169454, since the observed 
rotational temperature in this direction is within the typical dispersion of 
the sample. Here, again, we see a need for improved detailed models, in this 
case, of the excitation processes in diffuse molecular clouds. The precise set 
of observational results presented in this work would offer robust constraints 
for any such modeling efforts.

\section{CONCLUDING REMARKS}
Isotopologic ratios in interstellar molecules containing carbon are effective 
probes of the chemical processes active in diffuse environments. In this 
investigation, we examined optical absorption lines of CN and CH$^+$ along a 
total of thirteen lines of sight through diffuse molecular clouds. The very 
high signal-to-noise ratio observations enabled us to extract precise 
$^{12}$CN/$^{13}$CN and $^{12}$CH$^+$/$^{13}$CH$^+$ ratios, which were used to 
assess various predictions of diffuse cloud chemistry.

Our results on $^{12}$CH$^+$/$^{13}$CH$^+$ confirm that this ratio does not 
deviate from the ambient $^{12}$C/$^{13}$C ratio in local interstellar clouds, 
as expected if CH$^+$ is formed via nonthermal processes. Each of the McDonald 
sight lines in our sample with detectable absorption from $^{13}$CH$^+$ had at 
one time been found to have either a much lower or a much higher value of 
$^{12}$CH$^+$/$^{13}$CH$^+$. Such values are not confirmed here. In particular, 
the $^{12}$CH$^+$/$^{13}$CH$^+$ ratios that we derive toward the Pleiades stars, 
20~Tau and 23~Tau, are both indistinguishable from 70, precluding a 
$^{12}$C/$^{13}$C ratio near 40, as advocated by Hawkins \& Jura (1987), for the 
gas associated with this nearby cluster. Considering these results, we find no 
evidence for variation in $^{12}$C/$^{13}$C within the solar neighborhood. In 
contrast, the studies by Casassus et al. (2005) and Stahl et al. (2008), based 
on UVES spectra, found variation in $^{12}$CH$^+$/$^{13}$CH$^+$ at about the 
16\% level. These authors attributed the variation to intrinsic scatter in the 
local $^{12}$C/$^{13}$C ratio due to inefficient or incomplete mixing in the 
interstellar medium. From our analysis of some of the same UVES data, we 
conclude that the variation in $^{12}$CH$^+$/$^{13}$CH$^+$ is likely related to 
the difficulties of obtaining accurate $^{12}$CH$^+$/$^{13}$CH$^+$ ratios from 
these spectra. The major difficulties involve the placement of the rather 
poorly-defined continua, and the blending of absorption from $^{13}$CH$^+$ with 
weak $^{12}$CH$^+$ components blueward of the main components. It is probably 
not a coincidence that the $^{12}$CH$^+$/$^{13}$CH$^+$ ratios derived from our 
high-resolution McDonald spectra, which have flat, well-behaved continua, along 
lines of sight with simple CH$^+$ absorption profiles show little variation. 
While intrinsic scatter in the local $^{12}$C/$^{13}$C ratio may exist at some 
level, due to small differences in the amount of chemical enrichment and 
subsequent mixing from one site to another, it is probably too early to say 
that this variation has been detected.

Unlike in the case of CH$^+$, the isotopologic ratios in CN and CO show 
evidence for significant fractionation away from the ambient value of 
$^{12}$C/$^{13}$C in some diffuse environments. We attribute these results to 
the competing influences of isotope-selective photodissociation and isotopic 
charge exchange reactions, which vary with the strength of the UV radiation 
field and the physical conditions within a cloud. The results on 
$^{12}$CN/$^{13}$CN and $^{12}$CO/$^{13}$CO in lines of sight where both ratios 
have been determined are suggestive of the inverse relationship between these 
quantities that is anticipated, assuming that CN and CO coexist in diffuse 
cloud cores. This relationship was first directly suggested by Sheffer et al. 
(2007), but the isotopologic results for CN that existed at the time were not 
extensive enough to adequately test the idea. With the results presented here, 
the situation is now significantly improved, but further observations of both 
$^{12}$CN/$^{13}$CN and $^{12}$CO/$^{13}$CO would help to clarify the precise 
relationship between the two quantities. Specifically, measurements of 
$^{12}$CO/$^{13}$CO toward HD~169454, where $^{12}$CN/$^{13}$CN is 
unfractionated, and toward HD~170740, where CN is fractionated upward (by a 
factor of $1.9\pm0.5$ over the ambient value), would prove useful. One aspect 
of the chemistry of diffuse clouds that future modeling efforts will need to 
examine is the role played by isotopic charge exchange with CN, which we have 
found to be important in cold diffuse cloud cores. An ICE reaction involving CN 
is the most logical explanation for why the enhancements in $^{12}$CN/$^{13}$CN 
are not as large as expected along lines of sight with low $^{12}$CO/$^{13}$CO 
ratios, such as HD~73882 and HD~154368.

The effects of photochemical fractionation in carbon-bearing molecules are 
predicted to be the strongest in diffuse molecular clouds because dark clouds 
are better shielded from UV radiation and lack a significant abundance of 
carbon ions. This has been convincingly demonstrated by millimeter-wave 
emission studies of CN, CO, and H$_2$CO (see Milam et al. 2005). However, it is 
interesting to note that the expression developed in \S{} 4.3, which yields the 
fraction of carbon locked up in CO, may still apply in denser regions. Keene et 
al. (1998) detected 809 GHz emission from both $^{12}$C$^0$ and $^{13}$C$^0$ 
near the center of the Orion Nebula, finding a $^{12}$C$^0$/$^{13}$C$^0$ ratio 
of $58\pm12$. This was after Boreiko \& Betz (1996) observed 158 $\mu$m 
emission from both isotopic variants of C$^+$ in this region and found a 
$^{12}$C$^+$/$^{13}$C$^+$ ratio of $58\pm6$. The $^{12}$C$^{18}$O/$^{13}$C$^{18}$O 
ratio, also derived by Keene et al. (1998) from observations of the 2$-$1 and 
3$-$2 transitions, is $75\pm9$. While all of these ratios are consistent with 
an ambient $^{12}$C/$^{13}$C ratio of 70, the exact correspondence between the 
ratios in neutral and ionized carbon suggests that there is a real difference 
between those ratios and that found in C$^{18}$O. Since these observations 
sample gas in a well-known photon-dominated region (PDR), it is reasonable to 
assume that the slight increase in the $^{12}$C$^{18}$O/$^{13}$C$^{18}$O ratio 
over the ambient value is the result of selective photodissociation. An 
application of equation (2) to these measurements, with the substitution of 
$f$($^{13}$C$^+$) for $f$($^{13}$CN), would yield a value of $f$(CO) = 0.76, 
which is consistent with expectations for gas in a dense molecular cloud like 
that in Orion. We contrast this result with the values of $f$(CO) we obtain 
toward the diffuse molecular clouds in Ophiuchus, namely 0.40 toward 
$\rho$~Oph~A and 0.45 toward $\zeta$~Oph. The Ophiuchus results are appropriate 
for sight lines that probe the transition region between diffuse and dense 
clouds, where the conversion of C$^+$ into CO is expected to occur.

Beyond isotopologic ratios, the high quality optical spectra examined in this 
investigation allowed us to obtain precise rotational excitation temperatures 
for both $^{12}$CN and $^{13}$CN. Our weighted mean value of $T_{01}$($^{12}$CN) 
= $2.754\pm0.002$ K implies an excess over the temperature of the cosmic 
microwave background of only $29\pm3$~mK. This excess is considerably smaller 
than that suggested in the recent survey by S\l{}yk et al. (2008), but is also 
reduced compared to the excess found by Palazzi et al. (1992). Both of these 
previous studies indicated the need for an additional excitation mechanism 
beyond electron and neutral collisions (and the CMB) to account for the 
observed excitation of CN. The more modest excess found here eliminates the 
need for such a mechanism in the general ISM. Only three sight lines in our 
sample exhibit significant excesses in $T_{01}$ over the cosmic microwave 
background temperature. For the clouds in these directions, we derived upper 
limits to the electron densities in the range 0.3$-$0.9 cm$^{-3}$. The 
rotational excitation temperatures observed in $^{13}$CN, from strong 
detections of the $R$(0) and $R$(1) lines in four directions and the first 
interstellar detection of the $P$(1) line, show no excess over the temperature 
of the CMB.

\acknowledgments
We are grateful to Yaron Sheffer for allowing us the use of his program ISMOD 
and to Dan Welty for suggesting that we include the VLT sight lines in our 
analysis. D. L. L. thanks the Robert A. Welch Foundation for support through 
grant F-634. This research made use of the ESO Science Archive Facility as well 
as the SIMBAD database operated at CDS, France.


\begin{figure}[!t]
\centering
\includegraphics[width=0.45\textwidth]{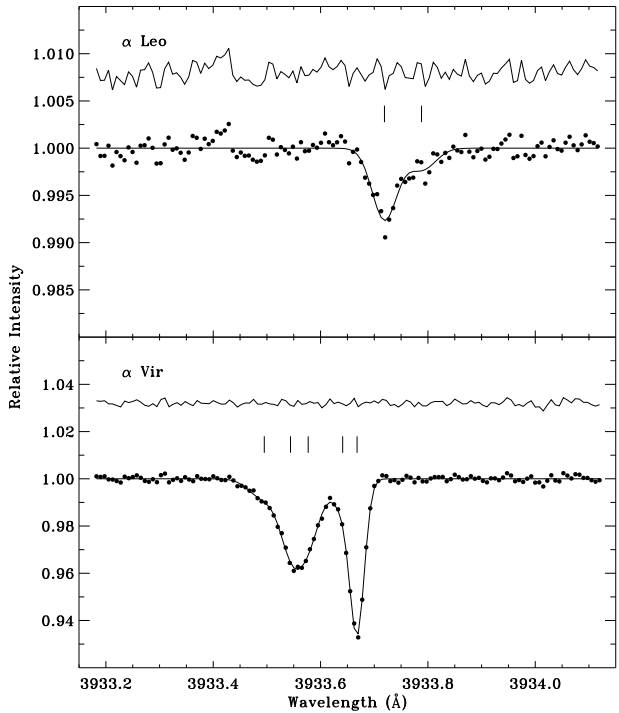}
\caption[Profile synthesis fits to the Ca~{\scriptsize II}~K lines toward 
$\alpha$~Leo and $\alpha$ Vir.]{Synthetic fits to the Ca~{\scriptsize II}~K 
profiles toward the unreddened stars $\alpha$ Leo (\emph{upper panel}) and 
$\alpha$ Vir (\emph{lower panel}). Plotting symbols are the same as in 
Figure~3. The Ca~{\scriptsize II} profile in the direction of $\alpha$ Leo is 
synthesized using two line-of-sight components, with LSR velocities of +4.4 and 
+9.7 km s$^{-1}$. For $\alpha$ Vir, five components comprise the fit. These 
have LSR velocities of $-$12.7, $-$8.9, $-$6.4, $-$1.5, and +0.5 km s$^{-1}$.}
\end{figure}

\begin{center}
APPENDIX\\[0.1in]INTERSTELLAR ABSORPTION TOWARD UNREDDENED STARS
\end{center}

In this appendix, we provide results on weak Ca~{\small I} ($\lambda$4226) and 
Ca~{\small II}~K ($\lambda$3933) absorption toward the stars $\alpha$ Leo and 
$\alpha$ Vir, which served as unreddened standard stars for our McDonald 
observations. Although $\alpha$ Lyr was also observed once as a standard, 
strong stellar lines in the vicinity of Ca~{\small I} and Ca~{\small II}~K, as 
well as lower S/N, prevent any interstellar features from being discerned in 
this direction. Table 10 presents the equivalent widths of Ca~{\small I} and 
Ca~{\small II}~K for the other two stars along with previous measurements and 
upper limits from the literature. Figure 21 shows our fits to the observed 
Ca~{\small II}~K profiles. For $\alpha$ Leo, the fit results in a total column 
density of $N$(Ca~{\small II}) = $(0.69\pm0.08)\times10^{10}$ cm$^{-2}$. A 
similar fit to the Ca~{\small I} profile yields a column density of 
$N$(Ca~{\small I}) = $(0.42\pm0.11)\times10^9$ cm$^{-2}$. For $\alpha$ Vir, our 
Ca~{\small II}~K profile synthesis fit gives $N$(Ca~{\small II}) = 
$(7.11\pm0.15)\times10^{10}$ cm$^{-2}$. No Ca~{\small I} absorption is detected 
in the direction of $\alpha$ Vir, but we derive an upper limit on the column 
density of $N$(Ca~{\small I}) $\lesssim0.21\times10^9$ cm$^{-2}$.

\clearpage

\begin{deluxetable}{lcccccccccc}
\tablecolumns{11}
\tablewidth{0.9\textwidth}
\tabletypesize{\small}
\tablecaption{Stellar and Observational Data}
\tablehead{ \colhead{} & \colhead{} & \colhead{} & \colhead{$B$} & 
\colhead{$E$($B-V$)} & \colhead{} & \colhead{$l$} & \colhead{$b$} & 
\colhead{$d$\tablenotemark{a}} & \colhead{Exp. Time\tablenotemark{b}} & 
\colhead{} \\
\colhead{Star} & \colhead{Name} & \colhead{Sp. Type} & \colhead{(mag)} & 
\colhead{(mag)} & \colhead{Ref.} & \colhead{(deg)} & \colhead{(deg)} & 
\colhead{(pc)} & \colhead{(hr)} & \colhead{S/N\tablenotemark{b}}  }
\startdata
\multicolumn{11}{c}{McDonald Sight Lines} \\
\hline
HD~23408 & 20~Tau & B7 III & 3.81 & 0.07 & 1 & 166.17 & $-$23.51 & $110\pm13$ & 
15.5/14.5 & 610/1240 \\
HD~23480 & 23~Tau & B6 IV & 4.11 & 0.10 & 1 & 166.57 & $-$23.75 & $110\pm13$ & 
33.0/33.0 & 1180/1930 \\
HD~24398 & $\zeta$~Per & B1 Ib & 2.97 & 0.34 & 2 & 162.29 & $-$16.69 & 
$301\pm68$ & 6.6/6.4 & 810/1500 \\
HD~147933 & $\rho$~Oph~A & B2 V & 5.26 & 0.45 & 3 & 353.69 & +17.69 & 90 & 
43.9/42.4 & 1100/1600 \\
HD~149757 & $\zeta$~Oph & O9.5 V & 2.60 & 0.32 & 3 & 6.28 & +23.59 & $140\pm14$ 
& 7.5/7.5 & 650/1590 \\
HD~179406 & 20~Aql & B3 V & 5.45 & 0.33 & 4 & 28.23 & $-$8.31 & 160 & 22.3/22.8 
& 900/1230 \\
\hline
\multicolumn{11}{c}{VLT/UVES Sight Lines} \\
\hline
HD~73882 & NX Vel & O8.5 Vn & 7.58 & 0.70 & 4 & 260.18 & +0.64 & 820 & 1.7 & 
1320 \\
HD~152236 & $\zeta^1$ Sco & B1 Ia+pe & 5.18 & 0.68 & 4 & 343.03 & +0.87 & 600 & 
0.2 & 1090 \\
HD~154368 & V1074 Sco & O9.5 Iab & 6.57 & 0.78 & 4 & 349.97 & +3.22 & 1100 & 
2.7 & 1680 \\
HD~161056 & & B1.5 V & 6.68 & 0.59 & 2 & 18.67 & +11.58 & 280 & 3.1 & 1480 \\
HD~169454 & V430 Sct & B1 Ia & 7.39 & 1.08 & 3 & 17.54 & $-$0.67 & 750 & 6.4 & 
1990 \\
HD~170740 & & B2 V & 5.93 & 0.48 & 4 & 21.06 & $-$0.53 & $213\pm43$ & 2.2 & 
1590 \\
HD~210121 & & B9 V & 7.83 & 0.31 & 3 & 56.88 & $-$44.46 & $210\pm45$ & 1.3 & 
1160 \\
\enddata
\tablenotetext{a}{Distances with error bars are from $\geq$ 4-$\sigma$ 
\emph{Hipparcos} results. Other distances are derived by means of spectroscopic 
parallax.}
\tablenotetext{b}{Exposure time and signal-to-noise ratio per resolution 
element for spectral region near CN $\lambda$3874/CH$^+$ $\lambda$4232. For the 
VLT/UVES sight lines, the values refer to the CN $\lambda$3874 region.}
\tablerefs{(1) White et al. 2001; (2) Wegner 2003; (3) Valencic et al. 2004; 
(4) Rachford et al. 2009.}
\end{deluxetable}


\begin{deluxetable}{lccc}
\tablecolumns{4}
\tablewidth{0.4\textwidth}
\tablecaption{Comparison of Total Equivalent Widths for CH$^+$ $\lambda$4232}
\tablehead{ \colhead{} & \colhead{$W_{\lambda}$($^{12}$CH$^+$)} & 
\colhead{$W_{\lambda}$($^{13}$CH$^+$)} & \colhead{} \\
\colhead{Star} & \colhead{(m\AA)} & \colhead{(m\AA)} & \colhead{Ref.} }
\startdata
20~Tau & 24.93 $\pm$ 0.07 & 0.36 $\pm$ 0.07 & 1 \\
 & 23.3 $\pm$ 0.3 & 0.67 $\pm$ 0.15 & 2 \\
 & 23.5 & 0.57 $\pm$ 0.11 & 3 \\
\\
23~Tau & 15.19 $\pm$ 0.05 & 0.23 $\pm$ 0.05 & 1 \\
 & 14.94 $\pm$ 0.02 & 0.15 & 4 \\
 & 14.0 $\pm$ 0.2 & 0.36 $\pm$ 0.08 & 2 \\
\\
$\rho$~Oph~A & 12.50 $\pm$ 0.06 & 0.21 $\pm$ 0.06 & 1 \\
 & 12.25 $\pm$ 0.09 & 0.11 $\pm$ 0.05 & 5 \\
\\
$\zeta$~Oph & 23.00 $\pm$ 0.06 & 0.36 $\pm$ 0.06 & 1 \\
 & 23.42 $\pm$ 0.03 & 0.29 & 6 \\
 & \phn23.8 $\pm$ 0.09 & 0.45 $\pm$ 0.06 & 7 \\
 & 23.25 $\pm$ 0.07 & 0.41 $\pm$ 0.02 & 8 \\
 & 22.62 $\pm$ 0.04 & 0.34 $\pm$ 0.01 & 9 \\
 & 22.5 $\pm$ 0.2 & 0.54 $\pm$ 0.1\phn & 10 \\
 & 21.7 $\pm$ 0.2 & 0.59 $\pm$ 0.10 & 11 \\
 & 23.5 & 0.36 $\pm$ 0.06 & 3 \\
\enddata
\tablerefs{(1) This Work; (2) Hawkins \& Jura 1987; (3) Vanden Bout \& Snell 
1980; (4) Stahl et al. 2008; (5) Federman et al. 2003; (6) Casassus et al. 
2005; (7) Hawkins et al. 1993; (8) Crane et al. 1991; (9) Stahl et al. 1989; 
(10) Hawkins et al. 1989; (11) Hawkins et al. 1985.}
\end{deluxetable}


\begin{deluxetable}{lccccccc}
\tablecolumns{8}
\tablewidth{0.8\textwidth}
\tabletypesize{\scriptsize}
\tablecaption{Comparison of Total Equivalent Widths for CN $\lambda$3874}
\tablehead{ \colhead{} & \multicolumn{3}{c}{$^{12}$CN} & \colhead{} & 
\multicolumn{2}{c}{$^{13}$CN} & \colhead{} \\
\cline{2-4} \cline{6-7} \\
\colhead{} & \colhead{$W_{\lambda} R$(0)} & \colhead{$W_{\lambda} R$(1)} & 
\colhead{$W_{\lambda} P$(1)} & \colhead{} & \colhead{$W_{\lambda} R$(0)} & 
\colhead{$W_{\lambda} R$(1)} & \colhead{} \\
\colhead{Star} & \colhead{(m\AA)} & \colhead{(m\AA)} & \colhead{(m\AA)} & 
\colhead{} & \colhead{(m\AA)} & \colhead{(m\AA)} & \colhead{Ref.} }
\startdata
\multicolumn{8}{c}{McDonald Sight Lines} \\
\hline
$\zeta$~Per & 8.88 $\pm$ 0.06 & 2.72 $\pm$ 0.06 & 1.36 $\pm$ 0.06 && 0.16 $\pm$ 
0.06 & \ldots & 1 \\
 & 8.70 $\pm$ 0.6\phn & 2.64 $\pm$ 0.2\phn & 1.00 $\pm$ 0.3\phn && \ldots & 
\ldots & 2 \\
 & 9.06 $\pm$ 0.17 & 2.79 $\pm$ 0.17 & 1.36 $\pm$ 0.17 && \ldots & \ldots & 3 
\\
 & \phn8.4 $\pm$ 0.21 & \ldots & \ldots && 0.12 $\pm$ 0.02 & \ldots & 4 \\
 & 8.99 $\pm$ 0.02 & 2.89 $\pm$ 0.02 & 1.30 $\pm$ 0.02 && \ldots & \ldots & 5 
\\
 & 8.06 $\pm$ 0.07 & 2.51 $\pm$ 0.04 & 1.28 $\pm$ 0.04 && \ldots & \ldots & 6 
\\
\\
$\rho$~Oph~A & 6.12 $\pm$ 0.04 & 1.95 $\pm$ 0.04 & 0.95 $\pm$ 0.04 && 0.15 
$\pm$ 0.04 & \ldots & 1 \\
 & 6.53 $\pm$ 0.4\phn & 2.18 $\pm$ 0.4\phn & 1.33 $\pm$ 0.4\phn && \ldots & 
\ldots & 2 \\
 & 6.1 $\pm$ 0.4 & 1.9 $\pm$ 0.3 & 1.1 $\pm$ 0.3 && \ldots & \ldots & 7 \\
 & 6.11 $\pm$ 0.48 & 2.08 $\pm$ 0.67 & 1.05 $\pm$ 0.59 && \ldots & \ldots & 8 
\\
 & 5.9 $\pm$ 0.3 & 2.5 $\pm$ 0.3 & 1.2 $\pm$ 0.7 && \ldots & \ldots & 9 \\
\\
$\zeta$~Oph & 7.98 $\pm$ 0.08 & 2.37 $\pm$ 0.08 & 1.27 $\pm$ 0.08 && 0.19 $\pm$ 
0.08 & \ldots & 1 \\
 & 8.31 $\pm$ 0.3\phn & 2.56 $\pm$ 0.4\phn & 1.29 $\pm$ 0.6\phn && \ldots & 
\ldots & 2 \\
 & 8.17 $\pm$ 0.04 & 2.50 $\pm$ 0.04 & 1.29 $\pm$ 0.04 && 0.27 $\pm$ 0.10 & 
\ldots & 3 \\
 & 8.16 $\pm$ 0.03 & \ldots & \ldots && 0.10 $\pm$ 0.05 & \ldots & 10 \\
 & 7.75 $\pm$ 0.04 & 2.45 $\pm$ 0.02 & 1.26 $\pm$ 0.02 && 0.19 $\pm$ 0.02 & 
\ldots & 11, 12 \\
 & 7.34 $\pm$ 0.12 & 2.13 $\pm$ 0.06 & 1.08 $\pm$ 0.07 && \ldots & \ldots & 6 
\\
 & 7.5 $\pm$ 0.3 & 2.6 $\pm$ 0.3 & 1.2 $\pm$ 0.7 && \ldots & \ldots & 9 \\
\\
20~Aql & 10.92 $\pm$ 0.07 & 3.29 $\pm$ 0.07 & 1.64 $\pm$ 0.07 && 0.20 $\pm$ 
0.07 & \ldots & 1 \\
 & 11.16 $\pm$ 0.2\phn & 3.99 $\pm$ 0.2\phn & 1.86 $\pm$ 0.2\phn && \ldots & 
\ldots & 2 \\
 & 11.5 $\pm$ 0.2 & 3.2 $\pm$ 0.2 & 1.6 $\pm$ 0.2 && \ldots & \ldots & 13 \\
\hline
\multicolumn{8}{c}{VLT/UVES Sight Lines} \\
\hline
HD~73882 & 34.01 $\pm$ 0.06 & 19.72 $\pm$ 0.06 & 12.07 $\pm$ 0.06 && 1.44 $\pm$ 
0.06 & 0.39 $\pm$ 0.06 & 1 \\
 & 34.83 $\pm$ 1.93 & 17.31 $\pm$ 0.55 & 10.23 $\pm$ 0.42 && \ldots & \ldots & 
8 \\
 & 31.8 $\pm$ 0.7 & 20.2 $\pm$ 1.3 & 13.7 $\pm$ 1.2 && 0.7 $\pm$ 0.5 & \ldots & 
14 \\
\\
HD~152236 (+6.0)\tablenotemark{a} & 7.09 $\pm$ 0.06 & 2.22 $\pm$ 0.06 & 1.20 
$\pm$ 0.06 && 0.13 $\pm$ 0.06 & \ldots & 1 \\
 & 6.09 $\pm$ 0.32 & 2.09 $\pm$ 0.60 & 0.74 $\pm$ 0.59 && \ldots & \ldots & 8 
\\
\\
HD~154368 (+5.2)\tablenotemark{a} & 25.30 $\pm$ 0.04 & 15.59 $\pm$ 0.04 & 9.66 
$\pm$ 0.04 && 1.13 $\pm$ 0.04 & 0.33 $\pm$ 0.04 & 1 \\
 & 26.58 $\pm$ 0.6\phn & 15.19 $\pm$ 0.7\phn & 10.01 $\pm$ 0.2\phn\phn && 
\ldots & \ldots & 2 \\
 & 25.34 $\pm$ 0.23 & 14.85 $\pm$ 0.22 & 9.21 $\pm$ 0.23 && \ldots & \ldots & 3 
\\
 & 22.5 $\pm$ 2.8 & 16.8 $\pm$ 2.3 & 8.1 $\pm$ 0.9 && \ldots & \ldots & 14 \\
 & 24.62 $\pm$ 0.13 & 15.12 $\pm$ 0.11 & 9.74 $\pm$ 0.09 && 0.82 $\pm$ 0.10 & 
0.25 $\pm$ 0.08 & 15 \\
\\
HD~161056 ($-$3.1) & 12.23 $\pm$ 0.05 & 4.49 $\pm$ 0.05 & 2.33 $\pm$ 0.05 && 
0.41 $\pm$ 0.05 & \ldots & 1 \\
 & 11.47 $\pm$ 0.6\phn & 3.38 $\pm$ 0.6\phn & \ldots && \ldots & \ldots & 2 \\
 & \phn8.7 $\pm$ 1.0 & 1.1 $\pm$ 1.0 & 2.0 $\pm$ 1.0 && \ldots & \ldots & 14 \\
\\
HD~161056 (+2.8) & 14.15 $\pm$ 0.06 & 4.26 $\pm$ 0.06 & 2.22 $\pm$ 0.06 && 0.43 
$\pm$ 0.06 & \ldots & 1 \\
 & 15.51 $\pm$ 0.6\phn & 4.54 $\pm$ 0.6\phn & \ldots && \ldots & \ldots & 2 \\
 & 15.6 $\pm$ 1.0 & 4.7 $\pm$ 1.0 & 3.0 $\pm$ 1.0 && \ldots & \ldots & 14 \\
\\
HD~169454 & 22.22 $\pm$ 0.04 & 16.50 $\pm$ 0.04 & 11.51 $\pm$ 0.04 && 1.92 
$\pm$ 0.04 & 0.47 $\pm$ 0.04 & 1 \\
 & 21.46 $\pm$ 0.2\phn & 15.75 $\pm$ 0.3\phn & 11.29 $\pm$ 0.2\phn && \ldots & 
\ldots & 2 \\
 & 25.10 $\pm$ 0.70 & 15.50 $\pm$ 0.88 & 10.85 $\pm$ 0.83 && \ldots & \ldots & 
8 \\
 & 22.2 $\pm$ 1.0 & 15.7 $\pm$ 0.4 & 11.7 $\pm$ 0.5 && 3.3 $\pm$ 1.3 & 1.0 
$\pm$ 0.5 & 14 \\
\\
HD~170740 & 13.53 $\pm$ 0.05 & 5.86 $\pm$ 0.05 & 3.27 $\pm$ 0.05 && 0.21 $\pm$ 
0.05 & \ldots & 1 \\
 & 13.60 $\pm$ 0.43 & 6.68 $\pm$ 0.61 & 3.82 $\pm$ 0.60 && \ldots & \ldots & 8 
\\
 & 14.0 $\pm$ 0.4 & 5.2 $\pm$ 0.8 & 2.6 $\pm$ 0.3 && \ldots & \ldots & 14 \\
\\
HD~210121 & 26.22 $\pm$ 0.07 & 12.20 $\pm$ 0.07 & 6.73 $\pm$ 0.07 && 0.94 $\pm$ 
0.07 & 0.33 $\pm$ 0.07 & 1 \\
 & 27.9 $\pm$ 1.8 & 10.5 $\pm$ 1.6 & 4.5 $\pm$ 1.0 && \ldots & \ldots & 16 \\
 & 26.8 $\pm$ 3.0 & 13.1 $\pm$ 2.6 & 8.1 $\pm$ 2.4 && \ldots & \ldots & 17 \\
 & 23.4 $\pm$ 2.0 & 10.0 $\pm$ 0.9 & 4.9 $\pm$ 0.9 && \ldots & \ldots & 14 \\
\enddata
\tablenotetext{a}{Equivalent widths listed for HD~152236 and HD~154368 refer to 
the dominant velocity component. For the $-$4.8 km~s$^{-1}$ component toward 
HD~152236, we find $W_{\lambda} R$(0) = $1.64\pm0.06$, $W_{\lambda} R$(1) = 
$0.39\pm0.06$, and $W_{\lambda} P$(1) = $0.14\pm0.06$, while for the $-$13.9 
km~s$^{-1}$ component toward HD~154368, we find $W_{\lambda} R$(0) = 
$0.28\pm0.04$ and $W_{\lambda} R$(1) = $0.09\pm0.04$.}
\tablerefs{(1) This Work; (2) Slyk et al. 2008; (3) Roth \& Meyer 1995; (4) 
Kaiser et al. 1991; (5) Kaiser \& Wright 1990; (6) Meyer \& Jura 1985; (7) Pan 
et al. 2004; (8) Palazzi et al. 1992; (9) Federman et al. 1984; (10) Hawkins et 
al. 1993; (11) Crane et al. 1989; (12) Crane \& Hegyi 1988; (13) Knauth et al. 
2001; (14) Gredel et al. 1991; (15) Palazzi et al. 1990; (16) Sheffer (private 
communication); (17) Welty \& Fowler 1992.}
\end{deluxetable}


\begin{deluxetable}{lcccc}
\tablecolumns{5}
\tablewidth{0.45\textwidth}
\tablecaption{Adopted Molecular Line Parameters}
\tablehead{ \colhead{} & \colhead{} & \colhead{$\lambda_{\mathrm{air}}$} & 
\colhead{} & \colhead{} \\
\colhead{Band} & \colhead{Line} & \colhead{(\AA)} & \colhead{$f$} & 
\colhead{Ref.} }
\startdata
CN $B$$-$$X$ (0,~0) & $R$(0) & 3874.608 & 0.0342\phn & 1 \\
 & $R$(1) & 3873.998 & 0.0228\phn & 1 \\
 & $P$(1) & 3875.763 & 0.0114\phn & 1 \\
 & $R$(2) & 3873.369 & 0.0205\phn & 1 \\
 & $P$(2) & 3876.310 & 0.0137\phn & 1 \\ \\
CN $B$$-$$X$ (1,~0) & $R$(0) & 3579.963 & 0.0030\phn & 1 \\ \\
CH$^+$ $A$$-$$X$ (0,~0) & $R$(0) & 4232.548 & 0.00545 & 2 \\ \\
CH$^+$ $A$$-$$X$ (1,~0) & $R$(0) & 3957.692 & 0.00331 & 2 \\
\enddata
\tablerefs{(1) Roth \& Meyer 1995; (2) Gredel et al. 1993.}
\end{deluxetable}


\begin{deluxetable}{lcccccc}
\tablecolumns{7}
\tablewidth{0.8\textwidth}
\tabletypesize{\small}
\tablecaption{CN Equivalent Width Ratios and Derived Column Densities and 
$b$-Values}
\tablehead{ \colhead{Star} & \colhead{(0,~0) $R$(1)/$P$(1)} & 
\colhead{$N$($N$=1)\tablenotemark{a}} & \colhead{$b$\tablenotemark{a}} & 
\colhead{(0,~0)/(1,~0) $R$(0)} & \colhead{$N$($N$=0)\tablenotemark{a}} & 
\colhead{$b$\tablenotemark{a}} }
\startdata
\multicolumn{7}{c}{McDonald Sight Lines} \\
\hline
$\zeta$~Per & 1.98 $\pm$ 0.11 & \ldots & $\infty$ & \ldots & \ldots & \ldots \\
$\rho$~Oph~A & 2.16 $\pm$ 0.10 & \ldots & $\infty$ & 11.33 $\pm$ 3.36 & \ldots 
& $\infty$ \\
$\zeta$~Oph & 1.74 $\pm$ 0.17 & \ldots & $\infty$ & 11.59 $\pm$ 1.54 & \ldots & 
$\infty$ \\
20~Aql & 2.01 $\pm$ 0.10 & \ldots & $\infty$ & \ldots & \ldots & \ldots \\
\hline
\multicolumn{7}{c}{VLT/UVES Sight Lines} \\
\hline
HD~73882 & 1.64 $\pm$ 0.01 & \phn9.84 $\pm$ 0.15 & 0.94 $\pm$ 0.03 & \ldots & 
\ldots & \ldots \\
HD~152236 (+6.0) & 1.91 $\pm$ 0.10 & \ldots & $\infty$ & 13.28 $\pm$ 3.26 & 
\ldots & $\infty$ \\
HD~154368 (+5.2) & 1.62 $\pm$ 0.01 & \phn7.91 $\pm$ 0.11 & 0.72 $\pm$ 0.02 & 
\phn4.65 $\pm$ 0.09 & 18.60 $\pm$ 0.47 & 0.68 $\pm$ 0.01 \\
HD~161056 ($-$3.1) & 1.86 $\pm$ 0.04 & \phn1.67 $\pm$ 0.10 & 0.52 $\pm$ 0.30 & 
15.64 $\pm$ 3.01 & \ldots & $\infty$ \\
HD~161056 (+2.8) & 2.00 $\pm$ 0.06 & \ldots & $\infty$ & 16.57 $\pm$ 3.08 & 
\ldots & $\infty$ \\
HD~169454 & 1.41 $\pm$ 0.01 & 11.93 $\pm$ 0.14 & 0.52 $\pm$ 0.01 & \phn2.83 
$\pm$ 0.05 & 32.06 $\pm$ 0.88 & 0.49 $\pm$ 0.01 \\
HD~170740 & 1.77 $\pm$ 0.03 & \phn2.42 $\pm$ 0.11 & 0.42 $\pm$ 0.08 & \phn7.18 
$\pm$ 0.54 & \phn5.86 $\pm$ 0.51 & 0.47 $\pm$ 0.04 \\
HD~210121 & 1.73 $\pm$ 0.02 & \phn5.44 $\pm$ 0.16 & 0.77 $\pm$ 0.08 & \phn5.94 
$\pm$ 0.16 & 14.38 $\pm$ 0.46 & 0.80 $\pm$ 0.02 \\
\enddata
\tablenotetext{a}{Column density (in $10^{12}$ cm$^{-2}$) and $b$-value (in 
km~s$^{-1}$) derived from the measured equivalent width ratio for a pair of CN 
transitions arising from the same rotational state (either $N$=1 or $N$=0).}
\end{deluxetable}


\begin{deluxetable}{lccccccc}
\tablecolumns{8}
\tablewidth{0.7\textwidth}
\tabletypesize{\scriptsize}
\tablecaption{CH$^+$ and CN Component Structure}
\tablehead{ \colhead{} & \multicolumn{3}{c}{CH$^+$} & \colhead{} & 
\multicolumn{3}{c}{CN} \\
\cline{2-4} \cline{6-8} \\
\colhead{} & \colhead{$v_{\mathrm{LSR}}$} & \colhead{$N$\tablenotemark{a}} & 
\colhead{$b$} & \colhead{} & \colhead{$v_{\mathrm{LSR}}$} & 
\colhead{$N$\tablenotemark{a}} & \colhead{$b$\tablenotemark{b}} \\
\colhead{Star} & \colhead{(km s$^{-1}$)} & \colhead{(10$^{12}$ cm$^{-2}$)} & 
\colhead{(km s$^{-1}$)} & \colhead{} & \colhead{(km s$^{-1}$)} & 
\colhead{(10$^{12}$ cm$^{-2}$)} & \colhead{(km s$^{-1}$)} }
\startdata
\multicolumn{8}{c}{McDonald Sight Lines} \\
\hline
20~Tau & +7.4 & 17.93 $\pm$ 0.09 & 1.8 && \ldots & \ldots & \ldots \\
 & +8.7 & 17.20 $\pm$ 0.09 & 1.8 && \ldots & \ldots & \ldots \\
\\
23~Tau & +7.0 & \phn1.06 $\pm$ 0.08 & 2.7 && \ldots & \ldots & \ldots \\
 & +10.0 & 18.42 $\pm$ 0.06 & 2.1 && \ldots & \ldots & \ldots \\
\\
$\zeta$~Per & +6.7 & 3.16 $\pm$ 0.08 & 2.2 && +6.6 & 2.31 $\pm$ 0.02 & 0.93 \\
\\
$\rho$~Oph~A & +1.5 & 6.95 $\pm$ 0.07 & 1.8 && +1.7 & 1.77 $\pm$ 0.01 & 0.42 \\
 & +3.2 & 8.89 $\pm$ 0.06 & 1.5 && \ldots & \ldots & \ldots \\
\\
$\zeta$~Oph & $-$0.9 & 31.86 $\pm$ 0.08 & 2.0 && $-$1.4 & 1.61 $\pm$ 0.02 & 
0.60 \\
 & \ldots & \ldots & \ldots && $-$0.2 & 0.45 $\pm$ 0.02 & 0.40 \\
\\
20~Aql & $-$6.6 & 0.50 $\pm$ 0.09 & 2.0 &&  \ldots & \ldots & \ldots \\
 & +2.6 & 3.15 $\pm$ 0.10 & 2.4 && +1.7 & 2.76 $\pm$ 0.02 & 1.34 \\
\hline
\multicolumn{8}{c}{VLT/UVES Sight Lines} \\
\hline
HD~73882 & +5.6 & 19.57 $\pm$ 0.09 & 2.8 && +5.7 & 27.66 $\pm$ 0.04 & 0.88 \\
 & +10.2 & \phn2.39 $\pm$ 0.07 & 1.9 && \ldots & \ldots & \ldots \\
\\
HD~152236 & $-$16.9 & \phn0.85 $\pm$ 0.15 & 3.1 && \ldots & \ldots & \ldots \\
 & $-$10.2 & \phn3.38 $\pm$ 0.12 & 2.0 && \ldots & \ldots & \ldots \\
 & $-$2.0 & 11.07 $\pm$ 0.15 & 3.1 && $-$4.8 & 0.38 $\pm$ 0.01 & 0.54 \\
 & +7.1 & \phn6.40 $\pm$ 0.14 & 3.0 && +6.0 & 2.21 $\pm$ 0.02 & 0.40 \\
\\
HD~154368 & $-$13.0 & \phn0.29 $\pm$ 0.06 & 1.9 && $-$13.9 & 0.06 $\pm$ 0.01 & 
0.20 \\
 & $-$6.8 & \phn0.90 $\pm$ 0.06 & 1.2 && \ldots & \ldots & \ldots \\
 & +2.4 & 14.44 $\pm$ 0.07 & 2.1 && \ldots & \ldots & \ldots \\
 & +6.6 & \phn7.29 $\pm$ 0.07 & 2.1 && +5.2 & 18.19 $\pm$ 0.03 & 0.68 $\pm$ 
0.01 \\
\\
HD~161056 & $-$7.9 & \phn0.51 $\pm$ 0.15 & 4.3 && \ldots & \ldots & \ldots \\
 & $-$3.2 & \phn0.77 $\pm$ 0.12 & 3.5 && $-$3.1 & 3.34 $\pm$ 0.01 & 1.02 \\
 & +2.4 & 24.31 $\pm$ 0.13 & 3.8 && +2.8 & 3.51 $\pm$ 0.02 & 2.00 \\
\\
HD~169454 & $-$3.0 & \phn4.41 $\pm$ 0.08 & 4.2 && \ldots & \ldots & \ldots \\
 & +5.5 & 14.96 $\pm$ 0.06 & 2.9 && +5.7 & 30.92 $\pm$ 0.05 & 0.50 $\pm$ 0.01 
\\
 & +15.4 & \phn1.16 $\pm$ 0.05 & 1.6 && \ldots & \ldots & \ldots \\
\\
HD~170740 & $-$1.1 & \phn0.51 $\pm$ 0.06 & 2.8 && \ldots & \ldots & \ldots \\
 & +5.1 & 16.46 $\pm$ 0.07 & 3.1 && +6.6 & 6.23 $\pm$ 0.02 & 0.46 $\pm$ 0.04 \\
\\
HD~210121 & $-$7.4 & 2.82 $\pm$ 0.10 & 1.5 && $-$6.7 & 14.33 $\pm$ 0.04 & 0.80 
$\pm$ 0.02 \\
 & $-$2.7 & 8.33 $\pm$ 0.13 & 2.6 && \ldots & \ldots & \ldots \\
\enddata
\tablenotetext{a}{Column densities refer to the $^{12}$C-bearing isotopologue.}
\tablenotetext{b}{We include 1-$\sigma$ errors on $b$-values derived by means 
of the doublet ratio method (see \S{} 3).}
\end{deluxetable}


\begin{deluxetable}{lccccccccc}
\tablecolumns{10}
\tablewidth{\textwidth}
\tabletypesize{\footnotesize}
\tablecaption{Total Molecular Column Densities and $^{12}$C/$^{13}$C Ratios}
\tablehead{ \colhead{} & \colhead{$N$($^{12}$CN)} & \colhead{} & 
\colhead{$\Delta v$(CN)} & \colhead{$N$($^{12}$CH$^+$)} & \colhead{} & 
\colhead{$\Delta v$(CH$^+$)} & \colhead{$N$($^{12}$CO)} & \colhead{} & 
\colhead{} \\
\colhead{Star} & \colhead{($10^{12}$ cm$^{-2}$)} & \colhead{$^{12}$CN/$^{13}$CN} 
& \colhead{(km s$^{-1}$)} & \colhead{($10^{12}$ cm$^{-2}$)} & 
\colhead{$^{12}$CH$^+$/$^{13}$CH$^+$} & \colhead{(km s$^{-1}$)} & 
\colhead{($10^{14}$ cm$^{-2}$)} & \colhead{$^{12}$CO/$^{13}$CO} & \colhead{Ref.} 
}
\startdata
\multicolumn{10}{c}{McDonald Sight Lines} \\
\hline
20~Tau & $\lesssim$ 0.06 & \ldots & \ldots & 35.13 $\pm$ 0.10 & 82.5 $\pm$ 16.7 
& $-$18.7 & \ldots & \ldots \\
23~Tau & $\lesssim$ 0.03 & \ldots & \ldots & 19.48 $\pm$ 0.06 & 72.7 $\pm$ 15.9 
& $-$18.6 & \ldots & \ldots \\
$\zeta$~Per & 3.28 $\pm$ 0.03 & 67.0 $\pm$ 28.0 & +13.2 & \phn3.16 $\pm$ 0.08 & 
\ldots & \ldots & 17.9 $\pm$ 0.5 & 108 $\pm$ 5\phn & 1 \\
$\rho$~Oph~A & 2.50 $\pm$ 0.02 & 54.0 $\pm$ 15.1 & +12.9 & 15.84 $\pm$ 0.08 & 
64.7 $\pm$ 18.5 & $-$19.0 & 19.2 $\pm$ 2.5 & 125 $\pm$ 36 & 1 \\
$\zeta$~Oph & 2.89 $\pm$ 0.03\tablenotemark{a} & 48.8 $\pm$ 19.5 & +13.0 & 
31.86 $\pm$ 0.08 & 75.2 $\pm$ 12.1 & $-$19.4 & 25.4 $\pm$ 1.6 & 167 $\pm$ 25 & 
1 \\
20~Aql & 3.94 $\pm$ 0.04 & 62.1 $\pm$ 22.2 & +12.8 & \phn3.64 $\pm$ 0.13 & 
\ldots & \ldots & 30$^{+25}_{-15}$ & \phn50 $\pm$ 15 & 2 \\
\hline
\multicolumn{10}{c}{VLT/UVES Sight Lines} \\
\hline
HD~73882 & 38.44 $\pm$ 0.06 & \phn85.0 $\pm$ 3.3\phn & +12.6 & 21.96 $\pm$ 0.11 
& \ldots & \ldots & 355 $\pm$ 208 & \phn25 $\pm$ 22 & 3 \\
HD~152236 & \phn3.52 $\pm$ 0.04\tablenotemark{a} & \phn64.9 $\pm$ 35.6 & 
+13.5\tablenotemark{b} & 21.70 $\pm$ 0.28 & \ldots & \ldots & \ldots & \ldots & 
\\
HD~154368 & 26.98 $\pm$ 0.04 & \phn70.7 $\pm$ 3.6\phn & +13.0 & 22.91 $\pm$ 
0.13 & \ldots & \ldots & 26.7 $\pm$ 5.5 & \phn37 $\pm$ 8\phn & 1 \\
HD~161056 & 10.10 $\pm$ 0.04 & \phn36.3 $\pm$ 3.5\phn & +14.0 & 25.59 $\pm$ 
0.24 & \ldots & \ldots & \ldots & \ldots & \\
HD~169454 & 44.73 $\pm$ 0.06 & \phn68.7 $\pm$ 1.3\phn & +12.8 & 20.52 $\pm$ 
0.12 & \ldots & \ldots & \ldots & \ldots & \\
HD~170740 & \phn8.77 $\pm$ 0.03 & 133.6 $\pm$ 33.0 & +13.4 & 16.97 $\pm$ 0.09 & 
\ldots & \ldots & \ldots & \ldots & \\
HD~210121 & 19.77 $\pm$ 0.05 & \phn68.4 $\pm$ 4.9\phn & +13.3 & 11.14 $\pm$ 
0.16 & \ldots & \ldots & 68 $\pm$ 14 & 102 $\pm$ 31 & 3 \\
\hline
Weighted mean && 67.5 $\pm$ 1.0 & +13.1 && 74.4 $\pm$ 7.6 & $-$18.9 \\
\enddata
\tablenotetext{a}{Total $^{12}$CN column densities for $\zeta$~Oph and 
HD~152236 do not include contributions from the $N=2$ level, but these 
contributions are likely to be small and within the uncertainties. If we assume 
$T_{12}=T_{01}$, we get $N$($N=2$) = $0.02\times10^{12}$ cm$^{-2}$ in both 
cases.}
\tablenotetext{b}{The velocity shift for CN $R$(0) toward HD~152236 was held 
fixed during profile synthesis, and thus is not included in the mean.}
\tablerefs{(1) Sheffer et al. 2007; (2) Hanson et al. 1992; (3) Sonnentrucker 
et al. 2007.} 
\end{deluxetable}


\begin{deluxetable}{lccccc}
\tablecolumns{6}
\tablewidth{0.6\textwidth}
\tablecaption{Comparison between Column Densities Derived from the (0,~0) and 
(1,~0) Bands}
\tablehead{ \colhead{} & \multicolumn{2}{c}{$N$(CH$^+$)\tablenotemark{a}} & 
\colhead{} & \multicolumn{2}{c}{$N$(CN)\tablenotemark{a}} \\
\cline{2-3} \cline{5-6} \\
\colhead{Star} & \colhead{(0,~0) $R$(0)} & \colhead{(1,~0) $R$(0)} && 
\colhead{(0,~0) $R$(0)} & \colhead{(1,~0) $R$(0)} }
\startdata
\multicolumn{6}{c}{McDonald Sight Lines} \\
\hline
20~Tau & 35.13 $\pm$ 0.10 & \ldots && \ldots & \ldots \\
23~Tau & 19.48 $\pm$ 0.06 & 20.16 $\pm$ 0.20 && \ldots & \ldots \\
$\zeta$~Per & \phn3.16 $\pm$ 0.08 & \ldots && \phn2.31 $\pm$ 0.02 & \ldots \\
$\rho$~Oph~A & 15.84 $\pm$ 0.08 & \ldots && \phn1.77 $\pm$ 0.01 & \phn1.53 
$\pm$ 0.48 \\
$\zeta$~Oph & 31.86 $\pm$ 0.08 & 33.24 $\pm$ 0.11 && \phn2.06 $\pm$ 0.02 & 
\phn2.13 $\pm$ 0.31 \\
20~Aql & \phn3.64 $\pm$ 0.13 & \ldots && \phn2.76 $\pm$ 0.02 & \ldots \\
\hline
\multicolumn{6}{c}{VLT/UVES Sight Lines} \\
\hline
HD~73882 & 21.96 $\pm$ 0.11 & 22.50 $\pm$ 0.26 && 27.66 $\pm$ 0.04 & \ldots \\
HD~152236 & 21.70 $\pm$ 0.28 & 21.42 $\pm$ 0.49 && \phn2.59 $\pm$ 0.03 & 
\phn2.36 $\pm$ 0.82 \\
HD~154368 & 22.91 $\pm$ 0.13 & 22.89 $\pm$ 0.24 && 18.25 $\pm$ 0.04 & 18.59 
$\pm$ 0.38 \\
HD~161056 & 25.59 $\pm$ 0.24 & 27.05 $\pm$ 0.42 && \phn6.85 $\pm$ 0.02 & 
\phn5.44 $\pm$ 0.84 \\
HD~169454 & 20.52 $\pm$ 0.12 & 21.14 $\pm$ 0.20 && 30.92 $\pm$ 0.05 & 30.78 
$\pm$ 0.52 \\
HD~170740 & 16.97 $\pm$ 0.09 & 17.62 $\pm$ 0.21 && \phn6.23 $\pm$ 0.02 & 
\phn5.64 $\pm$ 0.41 \\
HD~210121 & 11.14 $\pm$ 0.16 & 10.98 $\pm$ 0.40 && 14.33 $\pm$ 0.04 & 13.42 
$\pm$ 0.38 \\
\enddata
\tablenotetext{a}{Total (line-of-sight) column densities (in $10^{12}$ 
cm$^{-2}$) derived through profile synthesis of the (0,~0) and (1,~0) $R$(0) 
lines in CH$^+$ and CN.}
\end{deluxetable}


\begin{deluxetable}{lcccccc}
\tablecolumns{7}
\tablewidth{0.9\textwidth}
\tabletypesize{\small}
\tablecaption{CN Rotational Column Densities and Excitation Temperatures}
\tablehead{ \colhead{} & \colhead{$N$($N$=0)} & \colhead{$N$($N$=1)} & 
\colhead{$N$($N$=2)} & \colhead{$T_{01}$} & \colhead{$T_{12}$} & 
\colhead{$T_{01}$($^{13}$CN)} \\
\colhead{Star} & \colhead{($10^{12}$ cm$^{-2}$)} & \colhead{($10^{12}$ 
cm$^{-2}$)} & \colhead{($10^{12}$ cm$^{-2}$)} & \colhead{(K)} & \colhead{(K)} & 
\colhead{(K)} }
\startdata
\multicolumn{7}{c}{McDonald Sight Lines} \\
\hline
$\zeta$~Per & 2.31 $\pm$ 0.02 & 0.94 $\pm$ 0.02 & 0.04 $\pm$ 0.02 & 2.723 $\pm$ 
0.031 & 2.931 $\pm$ 0.411 & \ldots \\
$\rho$~Oph~A & 1.77 $\pm$ 0.01 & 0.68 $\pm$ 0.01 & 0.04 $\pm$ 0.01 & 2.657 
$\pm$ 0.026 & 3.136 $\pm$ 0.323 & \ldots \\
$\zeta$~Oph & 2.06 $\pm$ 0.02 & 0.82 $\pm$ 0.02 & \ldots & 2.702 $\pm$ 0.042 & 
\ldots & \ldots \\
20~Aql & 2.76 $\pm$ 0.02 & 1.12 $\pm$ 0.02 & 0.06 $\pm$ 0.02 & 2.728 $\pm$ 
0.028 & 3.155 $\pm$ 0.340 & \ldots \\
\hline
\multicolumn{7}{c}{VLT/UVES Sight Lines} \\
\hline
HD~73882 & 27.66 $\pm$ 0.04 & 10.48 $\pm$ 0.02 & 0.31 $\pm$ 0.02 & 2.631 $\pm$ 
0.004 & 2.693 $\pm$ 0.038 & 2.706 $\pm$ 0.200 \\
HD~152236 ($-$4.8) & \phn0.38 $\pm$ 0.01 & \phn0.12 $\pm$ 0.02 & \ldots & 2.454 
$\pm$ 0.178 & \ldots & \ldots \\
HD~152236 (+6.0) & \phn2.21 $\pm$ 0.02 & \phn0.81 $\pm$ 0.02 & \ldots & 2.588 
$\pm$ 0.033 & \ldots & \ldots \\
HD~154368 (+5.2) & 18.19 $\pm$ 0.03 & \phn8.40 $\pm$ 0.02 & 0.29 $\pm$ 0.01 & 
2.911 $\pm$ 0.004 & 2.815 $\pm$ 0.033 & 2.781 $\pm$ 0.192 \\
HD~161056 ($-$3.1) & \phn3.34 $\pm$ 0.01 & \phn1.59 $\pm$ 0.02 & 0.10 $\pm$ 
0.02 & 2.960 $\pm$ 0.018 & 3.333 $\pm$ 0.159 & \ldots \\
HD~161056 (+2.8) & \phn3.51 $\pm$ 0.02 & \phn1.46 $\pm$ 0.02 & 0.10 $\pm$ 0.02 
& 2.761 $\pm$ 0.020 & 3.383 $\pm$ 0.210 & \ldots \\
HD~169454 & 30.92 $\pm$ 0.05 & 13.30 $\pm$ 0.02 & 0.51 $\pm$ 0.01 & 2.804 $\pm$ 
0.003 & 2.881 $\pm$ 0.017 & 2.599 $\pm$ 0.082 \\
HD~170740 & \phn6.23 $\pm$ 0.02 & \phn2.44 $\pm$ 0.02 & 0.11 $\pm$ 0.02 & 2.672 
$\pm$ 0.011 & 2.989 $\pm$ 0.123 & \ldots \\
HD~210121 & 14.33 $\pm$ 0.04 & \phn5.31 $\pm$ 0.02 & 0.13 $\pm$ 0.02 & 2.604 
$\pm$ 0.007 & 2.589 $\pm$ 0.096 & 3.096 $\pm$ 0.383 \\
\hline
Weighted mean &&&& 2.754 $\pm$ 0.002 & 2.847 $\pm$ 0.014 & 2.652 $\pm$ 0.069 \\
\enddata
\end{deluxetable}


\begin{deluxetable}{lccc}
\tablecolumns{4}
\tablewidth{0.3\textwidth}
\tablecaption{Total Ca~{\scriptsize I} and Ca~{\scriptsize II} Equivalent 
Widths}
\tablehead{ \colhead{} & \colhead{$W_{\lambda}$(Ca~{\scriptsize I})} & 
\colhead{$W_{\lambda}$(Ca~{\scriptsize II})} & \colhead{} \\
\colhead{Star} & \colhead{(m\AA)} & \colhead{(m\AA)} & \colhead{Ref.} }
\startdata
$\alpha$ Leo & 0.12 $\pm$ 0.03 & 0.59 $\pm$ 0.07 & 1 \\
 & \ldots & $\lesssim$ 0.4 & 2 \\
\\
$\alpha$ Vir & $\lesssim$ 0.06 & 5.95 $\pm$ 0.12 & 1 \\
 & $\lesssim$ 0.2 & \ldots & 3 \\
 & \ldots & 5.4 & 4 \\
 & \ldots & 4.8 $\pm$ 0.1 & 2 \\
 & \ldots & 5.2 & 5 \\
 & \ldots & 5.7 & 6 \\
 & \ldots & 5 & 7 \\
\enddata
\tablerefs{(1) This Work; (2) Vallerga et al. 1993; (3) Welty et al. 2003; (4) 
Welty et al. 1996; (5) Lallement et al. 1986; (6) Hobbs 1978; (7) Hobbs 1975.} 
\end{deluxetable}

\end{document}